\documentclass[preprint,superscriptaddress,amsmath,amssymb,amsart,aps,prl]{revtex4-1}

\usepackage{graphicx}
\usepackage{dcolumn}
\usepackage{bm}
\usepackage{hyperref}
\usepackage{xcolor}
\usepackage{url}
\usepackage{xcolor}
\usepackage{tikz}
\usepackage{comment}
\usepackage{xr}
\usepackage{cleveref}
\usepackage{multibib}
\newcites{SM}{SM References}
\Crefname{figure}{Fig.}{Figs.}
\Crefname{equation}{Eq.}{Eqs.}

\definecolor{lime}{HTML}{A6CE39}
\DeclareRobustCommand{\orcidicon}{
	\begin{tikzpicture}
	\draw[lime, fill=lime] (0,0) 
	circle [radius=0.16] 
	node[white] {{\fontfamily{qag}\selectfont \tiny ID}};
	\draw[white, fill=white] (-0.0625,0.095) 
	circle [radius=0.007];
	\end{tikzpicture}
	\hspace{-2mm}
}
\foreach \x in {A, ..., Z}{\expandafter\xdef\csname orcid\x\endcsname{\noexpand\href{https://orcid.org/\csname orcidauthor\x\endcsname}
			{\noexpand\orcidicon}}
}

\newcolumntype{C}[1]{>{\centering\arraybackslash}p{#1}}

\usepackage{lineno}
\newcommand*{\WM }{The College of William \& Mary, Williamsburg, Virginia 23185, USA}

\newcommand*{\ANL }{Argonne National Laboratory, Lemont, Illinois 60439, USA}

\newcommand*{\MSU }{Mississippi State University, Mississippi State, Mississippi 39762, USA}

\newcommand*{\TEMP }{Temple University, Philadelphia, Pennsylvania 19122, USA}

\newcommand*{\JLAB }{Thomas Jefferson National Accelerator Facility, Newport News, Virginia 23606, USA}

\newcommand*{\BOULDER }{University of Colorado Boulder, Boulder, Colorado 80309, USA}

\newcommand*{\YER }{A.I. Alikhanyan  National  Science  Laboratory, Yerevan  Physics
Institute,  Yerevan  0036,  Armenia}

\newcommand*{\CUA }{Catholic University of America, Washington, DC 20064, USA}

\newcommand*{\REG }{University of Regina, Regina, Saskatchewan S4S 0A2, Canada}

\newcommand*{\ZAG }{University of Zagreb, Zagreb, Croatia}

\newcommand*{\HU }{Hampton University, Hampton, Virginia 23669, USA}

\newcommand*{\CNU }{Christopher Newport University, Newport News, Virginia 23606, USA}

\newcommand*{\UVA }{University of Virginia, Charlottesville, Virginia 22903, USA}

\newcommand*{\NCAT }{North Carolina A \& T State University, Greensboro, North Carolina 27411, USA}

\newcommand*{\UTENN }{University of Tennessee, Knoxville, Tennessee 37996, USA}

\newcommand*{\UCONN }{University of Connecticut, Storrs, Connecticut 06269, USA}

\newcommand*{\ODU }{Old Dominion University, Norfolk, Virginia 23529, USA}

\newcommand*{\OHIO }{Ohio University, Athens, Ohio 45701, USA}

\newcommand*{\UOY }{University of York, Heslington, York, YO10 5DD, UK}

\newcommand*{\FIU }{Florida International University, University Park, Florida 33199, USA}

\newcommand*{\JMU }{James Madison University, Harrisonburg, Virginia 22807, USA}

\newcommand*{\VM}{Virginia Military Institute, Lexington, Virginia 24450, USA}

\newcommand*{\SBU }{Stony Brook University, Stony Brook, New York 11794, USA}
\newcommand*{\TU}{Tsinghua University, Beijing 100084, China}

\begin{document}
\setcounter{secnumdepth}{2}

\title{Flavor, transverse momentum, and azimuthal dependence of charged pion multiplicities in semi-inclusive deep-inelastic scattering with 10.6 GeV electrons}




\author{P.\,Bosted}\affiliation{\WM}
\author{H.\,Bhatt}\affiliation{\MSU}
\author{S.\,Jia}\affiliation{\TEMP}
\author{W.\,Armstrong}\affiliation{\ANL} 
\author{D.\,Dutta}\affiliation{\MSU}
\author{R.\,Ent}\affiliation{\JLAB}   
\author{D.\,Gaskell}\affiliation{\JLAB} 
\author{E.\,Kinney}\affiliation{\BOULDER} 
\author{H.\,Mkrtchyan}\affiliation{\YER}

\author{S.\,Ali}\affiliation{\CUA}
\author{R.\,Ambrose}\affiliation{\REG} 
\author{D.\,Androic}\affiliation{\ZAG}  
\author{C.\,Ayerbe Gayoso}\affiliation{\MSU} 
\author{A.\,Bandari}\affiliation{\WM}   
\author{V.\,Berdnikov}\affiliation{\CUA}    
\author{D.\,Bhetuwal}\affiliation{\MSU}
\author{D.\,Biswas}\affiliation{\HU}  
\author{M.\, Boer}\affiliation{\TEMP}
\author{E.\,Brash}\affiliation{\CNU}    
\author{A.\,Camsonne}\affiliation{\JLAB}
\author{M.\,Cardona}\affiliation{\TEMP}          
\author{J.\,P.\,Chen}\affiliation{\JLAB}           
\author{J.\,Chen}\affiliation{\WM}    
\author{M.\,Chen}\affiliation{\UVA}             
\author{E.\,M.\,Christy}\affiliation{\HU}          
\author{S.\,Covrig}\affiliation{\JLAB}           
\author{S.\,Danagoulian}\affiliation{\NCAT}     
\author{M.\,Diefenthaler}\affiliation{\JLAB}     
\author{B.\,Duran}\affiliation{\TEMP}            
\author{C.\,Elliot}\affiliation{\UTENN}
\author{H.\,Fenker}\affiliation{\JLAB}           
\author{E.\,Fuchey}\affiliation{\UCONN}           
\author{J.\,O.\,Hansen}\affiliation{\JLAB}           
\author{F.\,Hauenstein}\affiliation{\ODU}       
\author{T.\,Horn}\affiliation{\CUA}             
\author{G.\,M.\,Huber}\affiliation{\REG}       
\author{M.\,K.\,Jones}\affiliation{\JLAB}          
\author{M.\,L.\,Kabir}\affiliation{\MSU}
\author{A.\,Karki}\affiliation{\MSU}            
\author{B.\,Karki}\affiliation{\OHIO} 
\author{S.\,J.\,D.\,Kay}\affiliation{\REG}\affiliation{\UOY}
\author{C.\,Keppel}\affiliation{\JLAB}           
\author{V.\,Kumar}\affiliation{\REG}
\author{N.\,Lashley-Colthirst}\affiliation{\HU}        
\author{W.\,B.\,Li}\affiliation{\WM}\affiliation{\MSU}               
\author{D.\,Mack}\affiliation{\JLAB}              
\author{S.\,Malace}\affiliation{\JLAB}           
\author{P.\,Markowitz}\affiliation{\FIU}       
\author{M.\,McCaughan}\affiliation{\JLAB}
\author{E.\,McClellan}\affiliation{\JLAB}
\author{D.\,Meekins}\affiliation{\JLAB}          
\author{R.\,Michaels}\affiliation{\JLAB}         
\author{A.\,Mkrtchyan}\affiliation{\YER}
\author{C.\,Morean}\affiliation{\UTENN}\affiliation{\JLAB}
\author{G.\,Niculescu}\affiliation{\JMU}        
\author{I.\,Niculescu}\affiliation{\JMU}        
\author{B.\,Pandey}\affiliation{\HU}\affiliation{\VM}           
\author{S.\,Park}\affiliation{\SBU}             
\author{E.\,Pooser}\affiliation{\JLAB}           
\author{B.\,Sawatzky}\affiliation{\JLAB}          
\author{G.\,R.\,Smith}\affiliation{\JLAB}             
\author{H.\,Szumila-Vance}\affiliation{\JLAB}\affiliation{\FIU}
\author{A.\,S.\,Tadepalli}\affiliation{\JLAB}
\author{V.\,Tadevosyan}\affiliation{\YER}        
\author{R.\,Trotta}\affiliation{\CUA}           
\author{H.\,Voskanyan}\affiliation{\YER}
\author{S.\,A.\,Wood}\affiliation{\JLAB}            
\author{Z.\, Ye}\affiliation{\ANL}\affiliation{\TU}
\author{C.\,Yero} \affiliation{\FIU}  
\author{X.\,Zheng}\affiliation{\UVA}        
\collaboration{for the Hall C SIDIS Collaboration}
\noaffiliation

\date{\today}

\begin{abstract}

Measurements of SIDIS multiplicities 
for $\pi^+$ and $\pi^-$ 
from proton and deuteron targets are reported on a grid of hadron kinematic variables
$z$, $P_{T}$, and $\phi^{*}$ for leptonic kinematic variables in the range $0.3<x<0.6$ and
$3<Q^2<5$ GeV$^2$. Data were acquired in 2018-2019 at Jefferson
Lab Hall C with a 10.6~GeV electron beam impinging on
10-cm-long liquid hydrogen and deuterium targets. Scattered
electrons  and charged pions were detected in the HMS and SHMS
spectrometers, respectively. The multiplicities were fitted for 
each bin in $(x,~Q^2,~z,~P_{t})$ to extract the 
$\phi^{*}$ - independent $M_0$ and the azimuthal modulations
$\langle \cos(\phi^{*}) \rangle$ and $\langle \cos(2\phi^{*}) \rangle$.
The $P_t$-dependence of the $M_0$ results was found to be remarkably consistent for the four cases 
studied: $ep\rightarrow e \pi^+ X$, 
$ep\rightarrow e \pi^- X$, $ed\rightarrow e \pi^+ X$, $ed\rightarrow e \pi^- X$
over the range $0<P_t<0.4$ GeV, as were the multiplicities evaluated near 
$\phi^* = 180^\circ$ over
the extended range $0<P_t<0.7$ GeV. The Gaussian widths of the $P_t$-dependence
exhibit a quadratic increase with $z$. 
The $\cos(\phi^{*})$
modulations were found to be 
consistent with zero for $\pi^+$, 
in agreement with previous world data, while the $\pi^-$ moments were, in many
cases, significantly greater than zero. 
The $\cos(2\phi^{*})$ modulations
were found to be consistent with zero. 
The higher statistical precision of this dataset of about 20,000 individual multiplicity
values, compared to previously published
data, should allow improved determinations of quark transverse momentum distributions 
and higher twist contributions.
\end{abstract}

      
\maketitle
\section{Introduction}
\begin{figure}[hbt!]
\centering
\includegraphics[width=\textwidth] {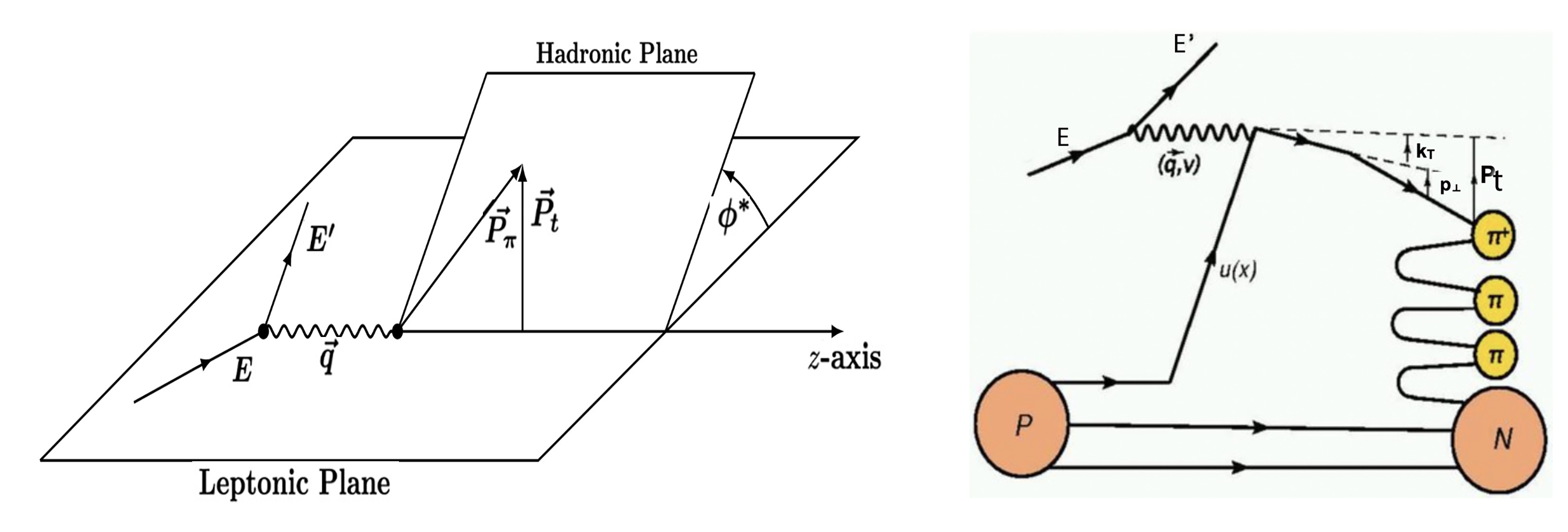}
 \caption{(left) Kinematics of the SIDIS process in the $\gamma^{*}P$ center of mass frame. (right) Simplified schematic of the pion SIDIS process, showing that the final transverse momentum of 
the leading pion, $P_{t}$,
arises from the convolution of the struck
quark's transverse momentum, $k_{T}$, with the
transverse momentum generated during the
fragmentation, $p_{\perp}$.}
 \label{fig:schematic1}
 \end{figure}

Over the last five decades, semi-inclusive deep-inelastic (SIDIS) lepton-nucleon scattering ($l N \rightarrow l^{'} h X$) has proven to be a key tool in building a more complete and accurate picture of the internal structure of the nucleon in terms of the partonic degrees of freedom of quantum chromodynamics (QCD). It has been instrumental in establishing that the collinear picture of the quark-parton model is incomplete. One of the most important advantages of SIDIS is the ability to measure the yield of hadrons ($h$) both in terms of the longitudinal momentum fraction $z$ and the transverse momentum $P_{t}$ (shown schematically in Fig.~\ref{fig:schematic1}). 
The SIDIS process in its simplest interpretation can be thought of as a
subset of deep-inelastic scattering (DIS), described by parton
distribution functions (PDFs), with a multiplicity function ($M$) that
indicates the probability of the DIS final-state containing a particular
meson with a particular momentum vector. 
In this highly simplified picture, the multiplicity dependence on $P_t$ 
arises from a convolution of the transverse momentum of the quark ($k_T$) and the transverse momentum
generated in the fragmentation process ($p_\perp$), in which the struck quark hadronizes
into multiple final-state particles. A comparison of SIDIS from protons ($u$-quark
dominated) and neutrons ($d$-quark dominated) could, in principle, be used to constrain the difference
between the average $k_T$ of up and down valence quarks in the nucleon.  Expanding the kinematic coverage  for both positive and
negative pions can help to distinguish differences in ``favored" and ``unfavored" fragmentation functions,
where ``favored" refers to a pion containing the struck quark. The azimuthal modulations of the
measured pion relative to the virtual photon direction are also sensitive to $k_T$, especially when
the incident electron or the target nucleon are polarized.~\cite{ansel05, ansel06}

In this paper, we present the results of a dedicated experiment at Jefferson Lab (JLab),
designed to augment the global SIDIS dataset through high precision measurements from
both hydrogen and deuteron targets, with measurements of both positively and negatively
charged pions in the kinematic region accessible with a 10.6 GeV electron beam and 
in-plane spectrometers. The high luminosity of this experiment has permitted 
binning the multiplicity results in a fine three-dimensional grid in 
$z$, $P_t$, and azimuthal angle $\phi^*$. Neither the beam nor the target
was polarized for this experiment. Nevertheless, our results provide a crucial
benchmark, at this energy, for the interpretation of SIDIS experiments with polarization
degrees of freedom.


\subsection{Formalism}
The semi-inclusive scattering of unpolarized electrons by unpolarized nucleons in the SIDIS kinematic region can be  described formally~\cite{Bacchetta07} in terms of structure functions (SF) as
\begin{eqnarray}
\label{eq:sidiscc}
\frac{d\sigma}{dx\,dy\,d\psi\,dz\,d\phi^{*}\,dP^2_{t}} &=&
\frac{\alpha^2}{xyQ^2}\frac{y^2}{2(1-\epsilon)}(1+\frac{\gamma^2}{2x})\\\nonumber
&&\Biggl\{F_{UU,T} +\epsilon F_{UU,L} + \sqrt{2\epsilon(1+\epsilon)}\cos(\phi^{*}) F^{\cos(\phi^{*})}_{UU}
+ \epsilon\cos(2\phi^{*}) F^{\cos(2\phi^{*})}_{UU} \Biggr\}
\end{eqnarray}
The four SF ($F_{UU,T},$ $F_{UU,L},$ $F^{\cos{(\phi^{*})}}_{UU},$ and $ F^{\cos{(2\phi^{*})}}_{UU}$) are all functions of $(x,Q^2,z,P_{t})$; $(x, Q^2, y)$ are the standard DIS virtual photon variables, $\epsilon$ is the
virtual photon polarization, and the detected hadron is characterized 
by its momentum fraction $z$, transverse momentum $P_t$, and azimuthal
angle $\phi^{*}$ of the hadronic reaction plane relative to the plane 
defined by the incident and scattered electron. We use the ``Trento" convention
for the definition of $\phi^*$~\cite{Bacchetta07}.
The fine structure constant is represented by $\alpha$  and 
the kinematic factor $\gamma = 2Mx/Q$, where $M$ is the nucleon mass. 
We define multiplicities as the ratio of the SIDIS cross section (Eq.~\ref{eq:sidiscc}) to the DIS cross section, calculated as a function of $(x, Q^2, y, \epsilon)$.


\subsection{Theoretical interpretation}
Although the intrinsic transverse momentum of the partons, $\vec{k}_{T}$, was already considered by Feynman when he introduced the parton model~\cite{FF1}, the initial experimental and theoretical focus was on the one-dimensional light-cone momentum variable $x$. In recent years significant advances have been made in incorporating 
$\vec{k}_{T}$ into the theoretical description of SIDIS processes. 
For example, the transverse momentum dependent (TMD) parton distribution
functions (PDF) and fragmentation functions (FF)~\cite{Muld96,Boer98} 
were introduced, and a TMD factorization formalism~\cite{Ji2004} was
developed. The factorization framework demonstrates that the hadron
transverse momentum arises from the transverse momentum of the quarks in
the nucleon, convoluted with the transverse momentum  generated during quark fragmentation. 
Both the TMD PDF and the FF depend on two independent variables: the TMD
on $x$ and $k_{T}$, while the FF depends on $z$ and the transverse momentum
$p_{\perp}$ of the hadron acquired during the fragmentation process. 
The TMD factorization was first shown for the high-energy limit (high values of the virtuality scale, $Q^2 >> \Lambda_{\text{QCD}}$) and moderate values of
$P_{t} \approx \Lambda_{\text{QCD}}$. However, its applicability at moderate $Q^2$ (2 - 4 GeV$^2$) 
has since been observed in several experiments~\cite{Tigran07,Hamlet08,Osipenko09,Asatur12}. 
Within this framework, and with the approximation that higher-order 
(higher-twist) corrections are suppressed by powers of $1/Q$, the SIDIS differential cross
section for polarized leptons on polarized nucleons is expressed in terms of 18 structure functions that are convolutions of various
TMD PDF and FF~\cite{Bacchetta07}. This large number of structure 
functions is a consequence of the fact that, for a spin-1/2 hadron, 
there are 8 TMD~\cite{Muld96,Bacchetta07,Ancel11}, each representing a unique correlation 
between the spin and the orbital motion of the partons. These TMD are 
parameterized using the world data on SIDIS and other 
processes~\cite{Barone10,Bacchetta11,Ancel14,Barone:2015ksa,Bacchetta17}.

As expected, the unpolarized SIDIS cross section can only provide information 
about the unpolarized TMD distribution functions
and the unpolarized TMD  fragmentation functions. 
The $\cos(\phi^{*})$ dependence was  predicted in 1978 by 
R.~Cahn~\cite{Cahn78} as a result of the interaction of the 
virtual photon with quarks in the nucleon possessing intrinsic 
transverse momentum. Both the  $\cos(\phi^{*})$  and 
$\cos(2\phi^{*})$ modulations receive contributions
from the Boer-Mulders effect \cite{BoerMuldersTimeR}, arising 
from a correlation between the quark's intrinsic transverse momentum 
and its transverse spin, coupled to the Collins 
fragmentation function~\cite{COLLINS1993161}, which preserves 
the correlation with fragmentation dependent on the struck quark's 
transverse spin. Phenomenological analyses by Barone et al.~\cite{Barone:2015ksa} 
stress that these structure functions are sensitive to higher-twist contributions.
Additionally,  
the transverse momentum dependence of the TMD and FF are expected 
to be approximately  Gaussian~\cite{ansel05}, for low values of 
$P_{t}$. To leading order, this simplification and momentum 
conservation give: $\langle\vec{P}^{2}_{t}\rangle \approxeq  \langle\vec{p}^{2}_{\perp}\rangle + z^{2} \langle\vec{k}^{2}_{T}\rangle$,   
 implying that the transverse momentum dependence of TMD and FF 
 can be parameterized by a normalized linear combination of a 
 Gaussian and a $z^2$-weighted Gaussian~\cite{Bacchetta17}. This 
 interpretation relies on $\langle\vec{p}^{2}_{\perp}\rangle$ being
 independent of $z$, which is manifestly incorrect as $z\rightarrow 0$ at finite values of $\nu$.

 \subsection{Previous experiments}
Some of the earliest SIDIS experimental studies in the 
valence quark region ($x>0.25$) were made at Cornell in the
1970s, using 12 GeV electrons~\cite{Cornell}, with an integrated
luminosity several orders of magnitude lower than the present
experiment with 10.6 GeV electrons. 
These experiments
demonstrated that multiplicities behave roughly as $(1-z)^2$ for
$z<0.7$, have an approximately Gaussian distribution in $P_t$, 
and have relatively small dependence on $\phi^*$ compared to exclusive
pion electroproduction. 
Subsequent experiments by the HERMES collaboration~\cite{HERMES:2012uyd,HERMES:2012kpt}
used 27 GeV electrons or positrons scattering from very thin targets
at DESY. The results are displayed as a function of $z$ averaged over
$(x,Q^2,P_t)$ and a function of $P_t$ averaged over $(x,Q^2,z)$, although
a later publication~\cite{Barone:2015ksa} displays azimuthal asymmetries averaged over $z$ in
some bins in $(x,Q^2)$. The HERMES results are all at a bit higher values of $W$
than the present experiment. 

Experiments at CERN by the EMC~\cite{EMC_PT_ref} and later by the COMPASS 
collaboration~\cite{COMPASS:2024gje,COMPASS_PT_ref} collaboration used 100-200 GeV muon beams
and focused on the sea quark region ($x<0.3$). The most recent
COMPASS publications show multiplicities as a function of $z$ averaged over $P_t$~\cite{COMPASS:2024gje},
and azimuthal dependencies as a function of $P_t$ averaged over $x$ and $z$~\cite{Benesova:2024cfk}.
Earlier publications~\cite{COMPASS_PT_ref} are for un-separated hadrons from a 
nuclear target (LiD). All of the CERN results are at considerably higher values
of $W$ than the present experiment. 

Global fits to the HERMES and COMPASS multiplicity results
have shown that spin-averaged cross sections can be reasonably well described as a convolution
of quark PDFs derived from DIS and Drell-Yan reactions 
with fragmentation functions (FF) derived from electron-positron colliders. 
They also demonstrated the usefulness of describing the production of the
leading meson that contains the struck quark flavor with ``favored" FF, while
other mesons are described by ``unfavored" FF, which exhibit a smaller strength
at high $z$ than favored FF. In this paper, we compare our multiplicity results to a very 
recent global fits to the HERMES and COMPASS multiplicity results
by the MAP collaboration~\cite{MAP24}, and to the the azimuthal 
asymmetries by Barone et al.~\cite{Barone:2015ksa}.

Jefferson Lab (JLab) offers the opportunity to perform much high luminosity
SIDIS experiments, due to beam currents of up to 70 $\mu$A, and an
effective duty factor of 100\%.
A pilot experiment at JLab Hall C with 5.5 GeV
electrons~\cite{Tigran07} showed an approximate duality between the 
results overlapping  the traditional nucleon resonance region, 
at low center-of-mass energy ($2<W<2.6$ GeV), and the results well above
the nucleon resonance region ($W>3$ GeV). This was the case as long as 
the electron-pion invariant
missing mass squared, $M_x^2$, was well above 2.5 GeV$^2$ 
(corresponding to $z<0.7$ at these
kinematic settings). A noticeable
peak centered at $M_x^2=1.5$ GeV$^2$ is likely due to the
semi-exclusive channel $e p \rightarrow e \pi \Delta(1232)$, which was
not subtracted in that analysis (but is subtracted in the present analysis). 
Simple phenomenological 
fits~\cite{Hamlet08,Asatur12} to these data attempted to disentangle 
the up and down valence quark $k_{T}$ widths, as well as favored and unfavored
FF widths, with the assumptions that the $\cos(\phi^*)$ dependence is
dominated by the Cahn term and that the fragmentation widths are 
independent of $z$ (both of which have since been shown to be incorrect).
Another JLab experiment using 5.5 GeV electrons was performed using the
low beam currents and large acceptance CLAS detector~\cite{Osipenko09}. 
Most of the results are
at relatively low $x$ and $Q^2$, where counting rates are the highest. The
results are shown as a function of $z$ averaged over $P_t$.

 
\section{The Experiment}
The experimental results prior to 2018 
cannot be considered conclusive in the quark valence region ($x>0.25$)
due to the limited kinematic coverage, low counting rates, inadequate
particle identification, and poor resolution in $\phi^{*}$ at low $P_{T}$. In order 
to overcome many of these limitations, a new experimental program was initiated 
in 2018 using beam energies up to 11 GeV with both the wide-acceptance, lower
luminosity CLAS12 detector in Hall B and the high-luminosity, small acceptance
spectrometers in Hall C. The broad program includes the use of beam and target
polarization, both light and heavy nuclear targets, a range of electron
beam energies, and identification of many final-state mesons.
In this paper, we report on spectrometer-based results for charged pions with an
unpolarized target at the highest available beam energy, from an experiment 
that was an integral part of the JLab SIDIS program and was completed in 2019. The experiment 
featured a wide range of $(x,Q^2)$ values (to study higher-twist contributions), full $\phi^{*}$ coverage for $P_{t}<0.3$ GeV, a larger $P_{t}$ range for $\phi^{*}$ near 180$^\circ$, and a broad range in $z$ (to 
help distinguish $k_{T}$ width from $p_{\perp}$ widths). The multiplicity results 
were published as a function of $W$ integrated over $\phi^*$ and $0<P_t<0.2$ GeV
in a recent Letter~\cite{Bhatt:2024prq}. In this paper, we present results in 
a fully 3-dinemsional grid in $(z,P_t,\phi^*$ for three values of $(x,Q^2$, spanning
$2.6<W<3.3$ GeV.

\begin{figure}[hbt!]
\centering

\includegraphics[width=0.6\textwidth] {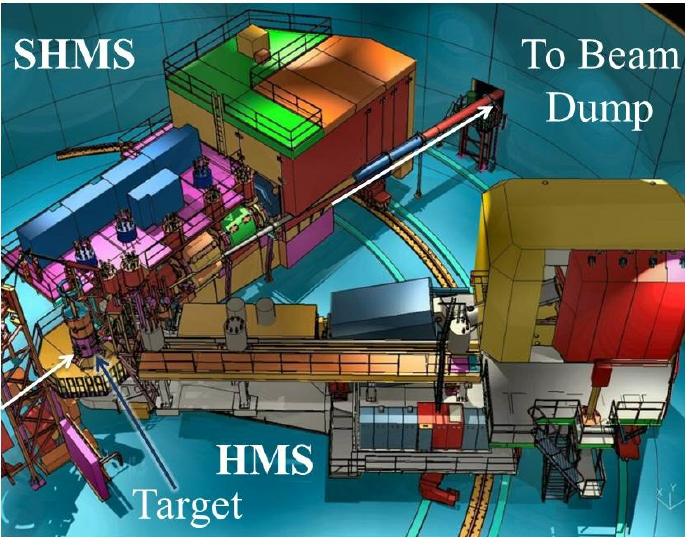}
 \caption{Overview of the experimental setup in Hall C at JLab.}
 \label{fig:hallc}
 \end{figure}

The experiment was carried out in spring 2018 and fall 2019, in 
Hall C at JLab. Electrons scattered from hydrogen and deuterium cryogenic
targets were detected in the High Momentum Spectrometer (HMS), with alternatively
positive and negative pions detected in the Super High Momentum Spectrometer (SHMS). Additional details about the experiment can be found in Refs.~\cite{Bhatt:2024prq,shms_nim}. An overview
of the experiment layout is shown in Fig.~\ref{fig:hallc}.

\subsection{Electron beam}
The experiment used a quasi-continuous-wave electron beam with beam 
energy of 10.6~GeV and beam currents ranging from 2 $\mu$A  to 70 $\mu$A. The spacing of
the beam micro-bunches was 4 ns for this experiment,
and the beam was rastered over a 2$ \times $2~mm$^2$ square pattern at $\approx$ 25 kHz. The beam 
energy was determined 
with a relative uncertainty of $<$0.05\%.  
A set of resonant-cavity-based 
beam-current monitors (BCM) was used to determine the total accumulated 
beam charge with a 
relative uncertainty of 
$\approx$ 0.5\%. 

\subsection{Targets}
The two main production targets were liquid hydrogen and liquid deuterium, each circulated
through 10 cm long and 3.4 cm radius aluminum cylinders.
At the pressure and temperature used in the experiment,
the  nominal areal density of the LH$_2$ was $714 \pm 14$ mg/cm$^2$ for kinematic settings I and III, and 
$718 \pm 8$ mg/cm$^2$ for kinematic setting II (the kinematic settings are listed in Table~\ref{tab:kin}). The nominal   areal density of the LD$_2$ was 
$1662 \pm 33$ mg/cm$^2$ for kinematic settings I and III, and 
$1662 \pm 17$ mg/cm$^2$ for kinematic setting II. 
A small reduction in the nominal density of 
the cryogenic targets due to beam heating was measured to be -0.023\%/$\mu$A. 
A so-called ``dummy target" consisting of two aluminum foils each with an areal density of 181  mg/cm$^2$
placed 10 cm apart was used to 
measure the contribution from the entrance and exit  end-caps of the 
cryogenic target cells.  The targets were cycled every few hours, reducing
the systematic errors on the ratio of multiplicities from hydrogen and deuterium,
compared to experiments in which targets are changed on a much longer time 
frame.

\begin{table*}[!hbt]
\caption{Beam energy $E$, HMS momentum $E^\prime$, HMS angle $\theta_{e}$, 
corresponding values of DIS variables $x$, $Q^2$, and $W$, and SHMS range of
central momentum ($p_{\pi}$) and angle ($\theta_{\pi}$) settings.} 
\label{tab:kin}\centering 
 \begin{tabular}{|c|c|c|c|c|c|c|c|c|c|}
\hline\hline
Setting & E & $E^\prime$& $\theta_{e}$ &$Q^{2}$ & $W$ & $x$ & $p_{\pi}$ & $\theta_{\pi}$ & run-period\\
\hline
 &(GeV) & (GeV) & (deg) &(GeV$^2$)  & (GeV)&  & (GeV) & (deg)& \\
\hline 
I   &10.6 &	5.240 &	13.50 &3.1 & 2.8 & 0.31 & 2.4 - 4.9 & 6.5 - 30 & Spring 2018\\
II  &10.6 &	3.307 &	19.70 &4.1 & 3.3 & 0.30 & 2.6 - 6.6& 6.5 - 22 & Spring 2018\\
III &10.6 &	5.240 &	16.30 &4.5 & 2.6 & 0.45 & 2.0 - 4.8 & 8 - 30& Fall 2019\\

\hline 
\hline
\end{tabular}
\end{table*}  
\subsection{Kinematics}  
The angle and momentum  of the electron 
arm (13$^{\circ}< \theta_e<$20$^{\circ}$, 3$<E^\prime<$5.2 GeV) 
and the hadron arm (6$^{\circ}< \theta_\pi<$30$^{\circ}$, 2$<P_\pi<$6 GeV) 
were chosen to map a region in $x$ and $z$ between 
0.25-0.65 and 0.3-0.7, respectively. The spectrometers are constrained 
to rotate around the target in a horizontal plane, which limits the out-of-plane 
angular coverage to about 0.08 radians. The angle, $\theta_{pq}$, between the 
electron three-momentum transfer, $\vec{q}$, and the hadron momentum 
was chosen to cover a range in $P_{t}$ up to 0.8 GeV. The electron kinematic settings
of the experiment are listed in Table~\ref{tab:kin}, along with the 
range of pion momenta and angles covered at each setting.

\subsection{Electron identification}
Scattered electrons were detected in the well-studied High Momentum
Spectrometer~\cite{hms_nim}, which has been in use since 1996.



Two drift chambers, each containing six planes of wires 
provided position and direction (track) information at the 
spectrometer focal plane with a resolution of $<$250~$\mu$m. 
The track information was then used to reconstruct the momentum 
and angle of the particle at  the target. 


A two-mirror threshold gas Cherenkov detector and a 
segmented Pb-glass calorimeter~\cite{cal_ref} were used to distinguish
electrons from pions (both of which generally passed the
cut on scintillator paddles). 
The Cherenkov detector gas mixture and pressure were set to
give a pion momentum threshold of 4.5 GeV. 
Detection efficiency was
determined to be $>99.5\%$ for settings I and II. 
For setting III, 
the average efficiency was 97.5\% for the electrons, independent of beam current.

\begin{figure}[hbt!]
\centering
\includegraphics[width=0.5\textwidth] {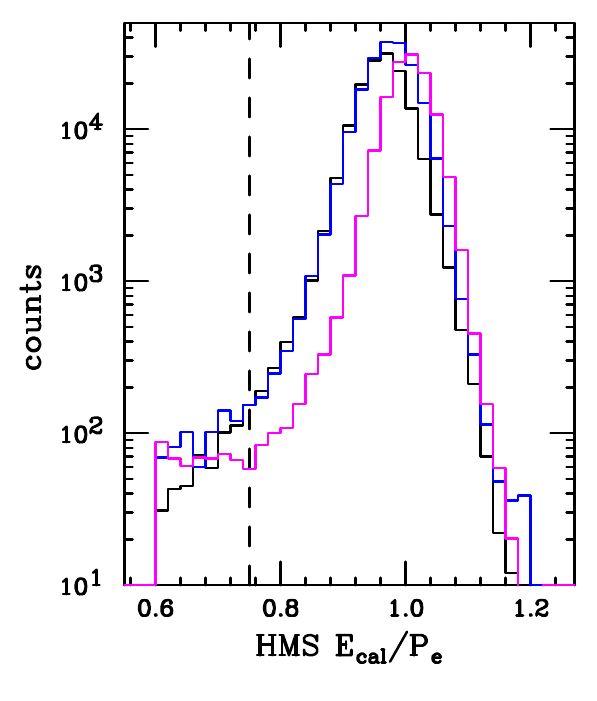}
 \caption{Accidental-subtracted distributions of normalized HMS calorimeter energy ($E_{cal}/p_e$) for kinematic setting I (black), II (blue), and III (magenta), for electron-pion coincidence events passing all cuts except the $E_{cal}/p_e$ $>$ 0.75 cut, which is indicated by the vertical dashed line.
 }
 \label{fig:shet}
 \end{figure}

Scattered electrons were identified in the segmented lead glass electromagnetic
calorimeter using the ratio of energy deposited in the blocks near the 
projected track position ($E_{cal}$)  to the track momentum ($P_e$). As 
illustrated in Fig.~\ref{fig:shet}, the $E_{cal}/P_e$  distributions
for each of the three kinematic settings show a narrow peak centered
on unity. 
The vertical dashed line at 0.75 shows the cut used
for electron identification. The electron detection efficiency of the HMS calorimeter was greater than
99.7\%. After correcting for accidental coincidences, the contamination
of pions in the final event sample was less than 0.5\%.

\subsection{Pion  identification}

Charged pions were detected in the Super High Momentum 
Spectrometer~\cite{shms_nim}, used for the first time in 2018. 
The momentum and angle ranges
used at each kinematic setting are listed in Table~\ref{tab:kin}, and were chosen 
to provide good coverage in the region $0.3<z<0.7$, along with as much coverage in
$P_t$ and $\phi^*$ as allowed by the spectrometer constraints.
The polarity of the spectrometer was alternated every few days in order to 
separately accumulate positively and negatively charged pions. This technique
provides identical acceptance for both charge states, resulting in small
systematic errors in the ratios of multiplicities, compared to large 
acceptance devices such as CLAS~\cite{CLAS6col, clas12col}. 

\begin{figure}[hbt!]
\centering
\includegraphics[width=0.5\textwidth] {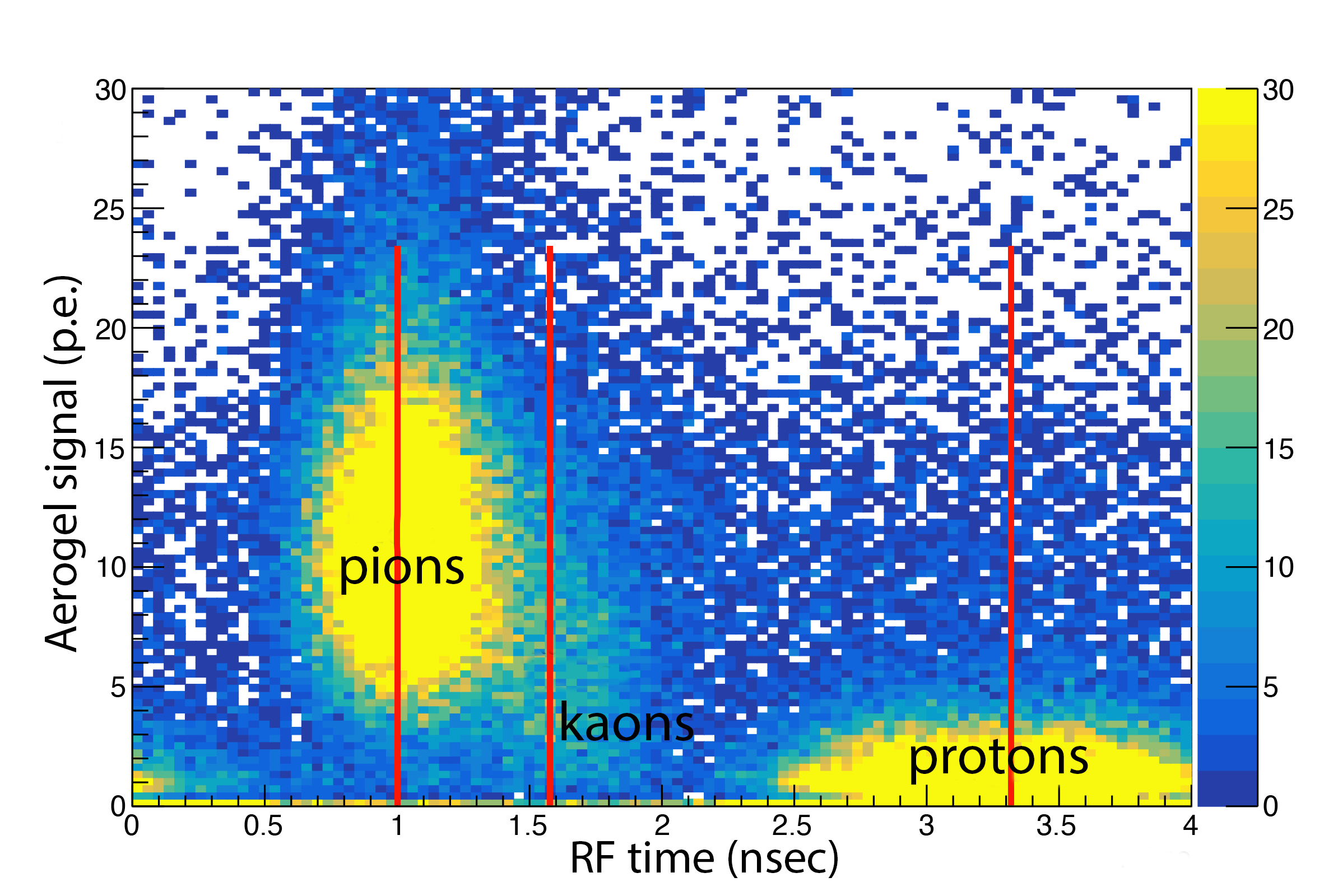}
 \caption{SHMS aerogel signals (in p.e.) from setting III, as a function of the arrival time of pions, kaons, and 
 protons relative to the beam micro-pulse time (RF time), modulo the 4 ns  bunch spacing, for
 particles with momenta $3.4<P_\pi<4.3$ GeV. The pion arrival time was shifted such that its peak is at 1~ns. The red vertical lines show the location of the pion, kaon, and proton peaks.}
 \label{fig:aerorf}
 \end{figure}

As in the HMS, the SHMS detector configuration 
included two pairs of segmented hodoscope 
planes (three scintillator planes and one quartz plane) separated by 2 m to provide fast timing signals
and rough particle trajectories. The resolution in particle speed was
sufficient to reject protons with momenta below 2 GeV. The average 
arrival time in the four paddles was compared to the arrival time of
the 4 ns  spaced beam micro-bunches-- dubbed as the radio-frequency (RF) time. With a flight
path of 22.5 m in the SHMS spectrometer, and a relative timing
cut of $\pm 0.7$ ns, it was possible to remove all protons and most
kaons from the event sample, as illustrated in Fig.~\ref{fig:aerorf}.
The efficiency of the timing cut was about 96\% for setting III. 
The RF timing signal was not operational for Settings I and II.


Two drift chambers, similar to those in the HMS, were used for
tracking. 
(
The tracking efficiency was found to
drop from  about 99.5\% at low rates to about 
97\% at the highest rates of particles entering the detector hut. 
To avoid pile-up effects in the tracking, the particle rate was kept
below 700 kHz by lowering the beam current to values as low as 2 $\mu A$.

To separate pions from electrons (or positrons), kaons,
and protons, three detectors were used: an aerogel Cherenkov detector, a heavy gas Cherenkov detector, and
an electromagnetic lead-glass calorimeter.
The aerogel detector was outfitted with multiple blocks with an index of refraction
of 1.015, corresponding to Cherenkov light thresholds of 0.9, 2.85, and 5.4 GeV for pions, 
kaons, and protons, respectively. Above threshold, an average of 10~p.e.~was produced.
 Below the Cherenkov threshold, kaons and protons often
produced a few p.e.~through knock-on scattering, as shown in Fig.~\ref{fig:aerorf}. Therefore, a minimum of 
4~p.e. was required for pion identification, with a corresponding efficiency of 95\%. 

The heavy gas Cherenkov detector contained C$_4$F$_{8}$O at less than 1 atm pressure, giving
a pion threshold of 2.61 GeV. It has 
a small inefficient region near the 
center of the detector for settings I and II, and a much larger region for setting III. 
Pions with momenta above 2.85 GeV were required to
have tracks outside the inefficient region and a light signal greater than 1 p.e. The
efficiency of this cut varied with momentum, increasing rapidly from 96\% at 2.85 GeV 
to $99\%$ for $P_\pi>3.2$ GeV.

\begin{figure*}[htb!]
\includegraphics[width=0.5\textwidth]{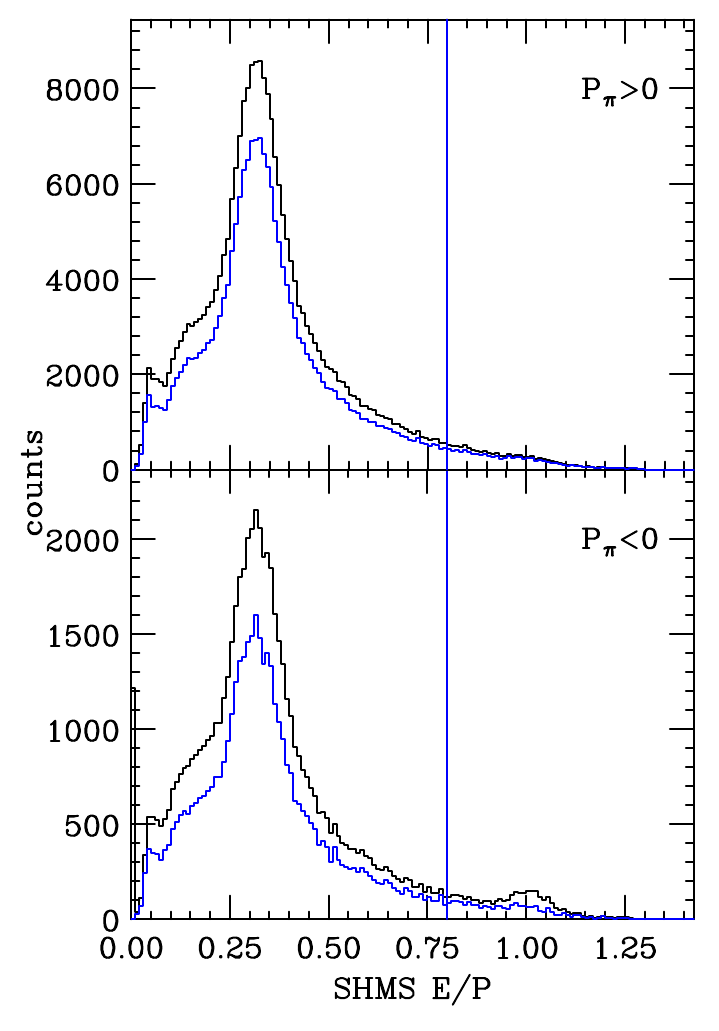} 
    \caption{Distributions of $E_{cal}/P_{\pi}$ in the SHMS calorimeter for positive (top panel) and
    negative (bottom panel) pion candidates. The sign of $P_{\pi}$ refers to the polarity of the spectrometer. The blue curves represent subsets of the black
    distributions with accidental coincidences subtracted. The vertical line at 0.8
    indicates the cut used to reject positrons (top panel) and electrons (bottom panel).}
    \label{fig:shtp}
\end{figure*}

The segmented lead-glass, 22 radiation-length electromagnetic calorimeter was used to separate hadrons from 
electrons. In contrast to the HMS, where electrons produced a narrow peak in $E_{cal}/P_e$ 
centered on unity, hadrons in the SHMS generally produced much less visible energy, as seen in the $E_{cal}/P_\pi$ distribution, because
the calorimeter is only about one hadronic interaction length in thickness. The distributions
in $E_{cal}/P_\pi$ are shown for both positive (top panel) and negative polarity (bottom panel) in 
Fig.~\ref{fig:shtp}, for good pion candidates selected by all cuts except that
on $E_{cal}/P_\pi$. A peak near unity can be seen in the negative polarity distribution, 
which we ascribe to accidental electron-electron coincidences. The peak is largely suppressed when accidental coincidences are removed (blue curves). There is essentially no evidence of electron-positron coincidences in the positive
polarity distributions. The residual distributions for $E_{cal}/P_\pi>0.8$ are likely 
dominated by charged-to-neutral pion conversions at the start of the hadronic shower process.
Nonetheless, we imposed a cut $E_{cal}/P_\pi<0.8$ to ensure no electron or positron 
contamination of the pion signal, with a typical efficiency of 0.94-0.97, depending
on the spectrometer momentum. 

The noble gas Cherenkov detector was installed only for settings I and II. With a pion
threshold of over 5 GeV, it was not directly used for pion identification. Its main use was
to provide a clean sample of electrons for calibrating the calorimeter. Additional information on the detectors used in the experiment, plots of the trigger efficiency, 
 and the detector efficiencies can be found in Ref.~\cite{Bhatt:2024prq, hem_thesis, shuo_thesis}.

\subsection{Electron-pion coincidence identification}
At the high luminosity of Hall C, there were many triggers for which an electron
and a pion originate from different beam bunches, spaced by 4 ns for this 
experiment. 
The average number of events in four accidental peaks (2 on either side of the true coincidence peak)  was subtracted from the central peak to select true electron-pion coincidences. For settings I and II (Spring 2018), the peak width was about 0.4 ns, consistent with the expected timing resolution of the trigger scintillators in both spectrometers, and a cut of $\pm 1$ ns was used to identify in-time coincidences. For setting III (Fall 2018), a mis-cabling problem caused the peak to be much wider (2.2 ns), reducing the ability to reject kaons and protons using coincidence timing. A wider cut of $\pm 2$ ns was therefore applied for setting III. Fortunately, the RF timing was operational for setting III, which more than compensated for this deficiency. The accidental-to-real ratio varied throughout the experiment from 10\% to 50\%.
\subsection{Readout trigger and data acquisition}
The trigger consisted of in-time signals on any three out of the four hodoscope
planes in each spectrometer. This ensured essentially 100\% trigger efficiency. 
The timing resolution of each plane was about
0.5 ns, resulting in an accuracy of typically 0.3 ns  for the electron-pion 
time difference. The trigger signaled the Data Acquisition (DAQ) system~\cite{coda} to read out pulse time and height information for all the detectors in both
spectrometers, and record them at rates of up to 3000 Hz. 

During the Spring 2018 runs (settings I and II), several problems caused a rate-dependent
loss of information for the desired electron-pion coincidences. 
The multiplicative 
correction factor $C_{DT}$
for these effects, determined by running with different beam currents under otherwise
identical conditions, was parameterized as:
$C_{DT}= 1.03 + 0.19 (R_{HMS} + R_{SHMS})$ where $R_{HMS}$ and  $R_{SHMS}$ are the trigger rates in the two spectrometers, in MHz. The factor of 1.03 at zero 
luminosity was obtained by comparing 
to Spring 2018 runs (setting III), for which most
of the problems of settings I and II were fixed, and
the luminosity correction is much smaller:
$C_{DT}= 1 + 0.04 (R_{HMS} + R_{SHMS})$.




 \subsection{Pair-symmetric background}
In inclusive electron scattering, an important background process
occurs when the measured electron originates from the decay
of a final-state hadron, especially for low values of $E^\prime/E$.
Approximately
equal contributions come from the Dalitz decay
$\pi^0 \rightarrow \gamma e^+ e^-$ and from the dominant 
$\pi^0 \rightarrow \gamma  \gamma$ decay, followed by subsequent
pair production from one of the photons in the target or spectrometer entrance window. This so-called pair-symmetric background is greatly reduced in SIDIS
compared to DIS by the requirement of a coincident pion at 
relatively large transverse momentum with respect to the 
electron beam. We performed a dedicated measurement of the
pair-symmetric background by reversing the polarity
of the HMS spectrometer and detecting the scattered positrons 
at two settings where the background was expected to be the largest. The results,
tabulated in Table~\ref{tab:pair}, indicate that the pair-symmetric
background is well below 0.5\%. 

\begin{table*}[!hbt]
\caption{Ratios of SIDIS rates with positrons 
compared to electrons in the HMS. The momentum of the HMS
was 3.6 GeV and the angle was 19 degrees. Pions were
measured in the SHMS with momenta of 2.6~GeV
and angles of 16 and 20 degrees. The sign of $P_p$ refers to the polarity of the spectrometer.}
\label{tab:pair}
\centering 
 \begin{tabular}{|c|c|c|c|c|}
\hline\hline
$P_p$($GeV$)& $\theta_p$($deg$) & target & $e^+/e^-$ \\
\hline
-2.6  &  16  &  p & $0.002\pm 0.002$ \\
      &      &  d & $0.004\pm 0.002$ \\
      &  20  &  d & $0.004\pm 0.002$ \\
      &      &  p & $0.004\pm 0.004$ \\
+2.6  &  20  &  p & $0.000\pm 0.002$ \\
      &      &  d & $0.002\pm 0.001$ \\
      &  16  &  d & $0.000\pm 0.001$ \\
\hline
\end{tabular}
\end{table*}
We also measured the pair-symmetric background for all
the momentum/angle settings of this experiment, by exchanging
the roles of the two spectrometers. 
The ratios of these 450  measurements lie in the
range 0 to 1\%, with an average of about 0.3\%.

Based on these results, we did not apply any pair-symmetric
correction, but assigned a systematic uncertainty of 0.3\% due
to this omission.



\section{Monte Carlo simulation}
A Monte Carlo (MC) simulation~\cite{simc}, named SIMC, was performed for each target and pair 
of spectrometer settings primarily to model the spectrometer acceptance and evaluate
radiative corrections. Another important use was to model pion and kaon decays (which 
led to muons and pions in the SHMS detector hut, respectively). For each setting,
the simulation was used to generate a large number of events for three distinct 
physics processes:  charged pion SIDIS itself (see Sec.~\ref{sec:sidis_model}); and 
the two background processes,
exclusive pion production and the semi-exclusive 
$\pi\Delta(1232)$ final-state arising from  bremsstrahlung emitted by either the incoming or the outgoing electron. These radiative contributions 
were treated in the angle peaking approximation using the formalism of Mo and 
Tsai~\cite{tsai}. The simulation also includes a detailed model of the targets, and geometrical 
acceptance and magnetic field map of the spectrometer magnets. The MC accounted for 
energy loss and multiple scattering in the target, vacuum windows, and detectors. 
Meson decays were allowed at any point along the particle trajectory, with the charged 
decay products tracked through the remainder of the spectrometer. The MC has been 
demonstrated to accurately reproduce the performance of the Hall C spectrometers~\cite{hem_thesis}. The multiplicity and cross section models used in the simulation are 
described in the next three sections. The SIDIS model was improved by scaling the ratio of measured yields to the MC yield and iterating this process. 
\subsection{SIDIS model}
\label{sec:sidis_model}
After two iterations, the
charged pion SIDIS cross section model, obtained 
using a global fit to our results, augmented with world data is given by:
\begin{equation}
\label{eq:mult}
\sigma_{SIDIS}=\sigma_{DIS}(x,Q^2,\epsilon)
M_{SIDIS}(z,P_t,\phi^{*},x,Q^2).
\end{equation}
The inclusive DIS cross section $\sigma_{DIS}(x,Q^2,\epsilon)$ is from a  global fit to
all world data available by the year 2020 for electrons
scattering from both proton and deuteron
targets~\cite{bosted_fit}. It is the most comprehensive model of the DIS cross section measured with 
the electron spectrometer used in this experiment, and the inclusive 
data collected at the kinematic settings used in this paper were 
found to be consistent with this model to within a few percent.

The $z$-dependence of the multiplicity function
$M_{SIDIS}(z,P_t,\phi^{*},x,Q^2)$ is given by:
\begin{eqnarray}
\nonumber
      zM_{p\pi^+}(z,x,Q^2) = (q_u^2 u    D_f + 
                 q_u^2 \bar{u} D_u +
       	  q_d^2 d    D_u + 
                 q_d^2 \bar{d} D_f + 
       	  q_s^2 s    D_u + 
                 q_s^2 \bar{s} D_u)/
                 \sum{(q_i)^2}  &&\\ \nonumber             
      zM_{p\pi^-}(z,x,Q^2) = (q_u^2 u    D_u + 
                 q_u^2 \bar{u} D_f +
       	  q_d^2 d    D_f + 
                 q_d^2 \bar{d} D_u + 
       	  q_s^2 s    D_u + 
                 q_s^2 \bar{s} D_u)/\sum{(q_i)^2}  &&\\ \nonumber               
      zM_{n\pi^+}(z,x,Q^2) = (q_u^2 d    D_f + 
                 q_u^2 \bar{d} D_u +
       	  q_d^2 u    D_u + 
                 q_d^2 \bar{u} D_f + 
       	  q_s^2 s    D_u + 
                 q_s^2 \bar{s} D_u)/
                 \sum{(q_i)^2}&&\\
      zM_{n\pi^-}(z,x,Q^2) = (q_u^2 d    D_u + 
                 q_u^2 \bar{d} D_f +
       	  q_d^2 u    D_f + 
                 q_d^2 \bar{u} D_u + 
       	  q_s^2 s    D_u + 
                 q_s^2 \bar{s} D_u)/
     \sum{(q_i)^2} &&
\end{eqnarray}     
where, $M_{p/n \pi^{\pm}}(z,x,Q^2)$ are the charged pion multiplicities from the proton (p) and neutron (n),  $q_i$ are the quark charges, the quark distribution
functions $u,d,s,\bar{u},\bar{d},\bar{s}$ were taken
from CTEQ5~\cite{cteq5}, and the favored and unfavored fragmentation
functions $D_f$ and $D_u$ were parameterized as:
\begin{equation}
D_{f/u} = p_{1} \zeta ^{(p_{2} + p_{4}s_v + p_{9}W^{-1})}  
           (1-\zeta)^{(p_{3} + p_{5}s_v + p_{10}W^{-1})} 
          (1 + p_{6}\zeta + p_{7}\zeta^2 + p_{8}\zeta^3) 
          (1 + p_{11}W^{-1} + p_{12}W^{-2}),
          \end{equation}        
          where, $s_v = \ln(Q^2/2)$ and the target mass corrections were applied using~\cite{Schienbein_2008}
          \begin{equation}
         \zeta = z \frac{1 + \sqrt{1 - 4 x^2(m_{\pi}^2 + P_t^2)/z^2Q^2}}{1 + \sqrt{1 + 4 x^2 M^2 / Q^2}}
         \end{equation}

The fit parameters $p_i$ were obtained from an iterative fit
to the data of this experiment, and are given in Table~\ref{tab:dpar}.
\begin{table}[hbt!]
\caption{Table of parameters used for $D_{f/u}$.}
\label{tab:dpar}
\centering 
 \begin{tabular}{|c|c|c|c|c|c|c|c|c|c|c|c|c|}
\hline\hline
 &$p_1$ & $p_2$ & $p_3$ & $p_4$ & $p_5$ & $p_6$ & $p_7$ & $p_8$ & $p_9$ & $p_{10}$ & $p_{11}$ & $p_{12}$ \\
\hline
$D_f$ & 1.0424 &  -0.1714 &   1.8960 &  -0.0307& 
  0.1636&  -0.1272& -4.2093 &  5.0103& 
  2.7406&  -0.5778&0   3.5292&   7.3910 \\
$D_u$ &  0.7840 &   0.2369 &   1.4238 &   0.1484 & 
  0.1518 &  -1.2923 &  -1.5710 &  3.0305 &
  1.1995 &   1.3553 &   2.5868 &   8.0666 \\
\hline
\end{tabular}
\end{table}

The $P_t$-dependence of the multiplicity
functions was incorporated as: 
\begin{equation}
M_{p/n\pi^{\pm}}(z,P_t,\phi^{*},x,Q^2) = \frac{1}{2\pi}M_0(z,x,Q^2)
b e^{-b P_t^2},
\end{equation}
i.e., a Gaussian distribution with the parameter 
$b=(0.12 z^2 + 0.2)^{-1}$ GeV$^{-2}$, common to all processes with amplitude $M_0(z,x,Q^2)$.
Note that we do not have any azimuthal dependence in
this fit, consistent with the $\pi^{+}$ results of the
present experiment. Also note that we do not have
a factorized expression: the multiplicity function
depends on the electron variables $(x,Q^2,W)$, which
we found necessary to describe the data of this
experiment. 

\subsection{Exclusive pion production model}
\label{sec:excl_model}
The cross section for exclusive charged pion 
electroproduction is defined as:
\begin{equation}
        \sigma=\frac{1.359}{(s - M^2)^2}(\sigma_T + 
        \epsilon  \sigma_L + 
        \epsilon  \cos(2\phi^{*})  \sigma_{TT} + 
        \sqrt{2 \epsilon (1+\epsilon})  
        \cos(\phi^{*}) \sigma_{LT})       
        \end{equation}
        where all relevant units are in GeV, $M$ is the average nucleon mass, and the longitudinal and transverse cross sections $\sigma_L,\sigma_T$, as well as two interference terms $\sigma_{LT}$ and $\sigma_{TT}$ are given in terms of the pion form factor $F_{\pi}$ by:
        \begin{eqnarray}
        \nonumber
         F_\pi &=&  
       (1 + p_1 Q^2 + p_2Q^4)^{-1} \\ \nonumber
      \sigma_L &=& (p_{3} + p_{15}/Q^2)  |t| / 
        (|t| + 0.02)^2  Q^2 F_{\pi}^2(s^{p_{11}} + \sqrt{s^{p_{17}}})e^{p_{4} |t|}  \\ \nonumber
       \sigma_T &=& p_{5} / Q^2 e^{p_{6}  Q^4}
        / (s^{p_{12}} + \sqrt{s^{p_{16}}})
       e^{p_{14}  |t|}  \\ \nonumber 
      \sigma_{LT} &=&(p_{7} / (1 + p_{10}  Q^2))
        e^{p_{8}  |t|}  \sin(\theta_{cm})
         / s^{p_{13}} \\ 
       \sigma_{TT} &=&(p_{9} / (1 + Q^2))  
         e^{-7.0 |t|}  \sin(\theta_{cm})^2 
       \end{eqnarray}
The parameters $p_i$ for exclusive pion production from the proton ($e p \rightarrow e \pi^+ n$) and the neutron ($e n \rightarrow e \pi^- p$) are obtained from fits to world data on LT separated pion electroproduction cross sections~\cite{th_thesis} and are shown in Tables~\ref{tab:pioncc1} and ~\ref{tab:pioncc2} . 

\begin{table}[hbt!]
\caption{Table of parameters used for exclusive pion electroproduction cross sections.}
\label{tab:pioncc1}
\centering 
 \begin{tabular}{|c|c|c|c|c|c|c|c|c|c|c|}
\hline\hline
 &$p_1$ & $p_2$ & $p_3$ & $p_4$ & $p_5$ & $p_6$ & $p_7$ & $p_8$ & $p_9$ & $p_{10}$ \\
\hline
$n_{\pi^- p}$ & 1.60077 &
          -0.01523 &
          37.08142 &
          -4.11060&
          23.26192&
           0.00983&
           0.87073&
          -5.77115&
        -271.08678&
           0.13766\\
$p_{\pi^+ n}$ & 1.75169&
           0.11144&
          47.35877&
          -4.69434&
           1.60552&
           0.00800&
           0.44194&
          -2.29188&
         -41.67194&
           0.69475\\
\hline
\end{tabular}
\end{table}
\begin{table}[hbt!]
\caption{Table of parameters used for exclusive pion electroproduction cross sections.}
\label{tab:pioncc2}
\centering 
 \begin{tabular}{|c|c|c|c|c|c|c|c|}
\hline\hline
 & $p_{11}$ & $p_{12}$ & $p_{13}$ & $p_{14}$ & $p_{15}$ & $p_{16}$ & $p_{17}$\\
\hline
$n_{\pi^- p}$ &-0.00855&
           0.27885&
          -1.13212&
          -1.50415&
          -6.34766&
           0.55769&
          -0.01709\\
$p_{\pi^+ n}$ &0.02527&
          -0.50178&
          -1.22825&
          -1.16878&
           5.75825&
          -1.00355&
           0.05055 \\
\hline
\end{tabular}
\end{table}

\subsection{Model for $\pi \Delta$}
\label{sec:delta_model}
We modeled the semi-exclusive reactions with
$\pi\Delta(1232)$ in the final-state by simply
scaling fully exclusive pion electroproduction
by the effective Clebsch–Gordan coefficients determined
from a fit to the data of this experiment.
The coefficients are given by: 
$$(e p \rightarrow e \pi^+ \Delta^0) /
  ( e p \rightarrow e \pi^+ n) = 0.4$$
$$(e n \rightarrow e \pi^+ \Delta^-) /
  ( e p \rightarrow e \pi^+ n) = 0.8$$
$$(e p \rightarrow e \pi^- \Delta^{++}) /
  (e n \rightarrow e \pi^- p) = 0.55$$
$$(e n \rightarrow e \pi^- \Delta^+) /
  (e n \rightarrow e \pi^- p) = 1.0$$

The final-state missing mass was simulated using a Breit-Wigner
distribution for the $\Delta(1232)$.

\subsection{Kinematic dependence of radiative corrections}
\begin{figure}[hbt!]
\centering
\includegraphics[width=0.8\textwidth] {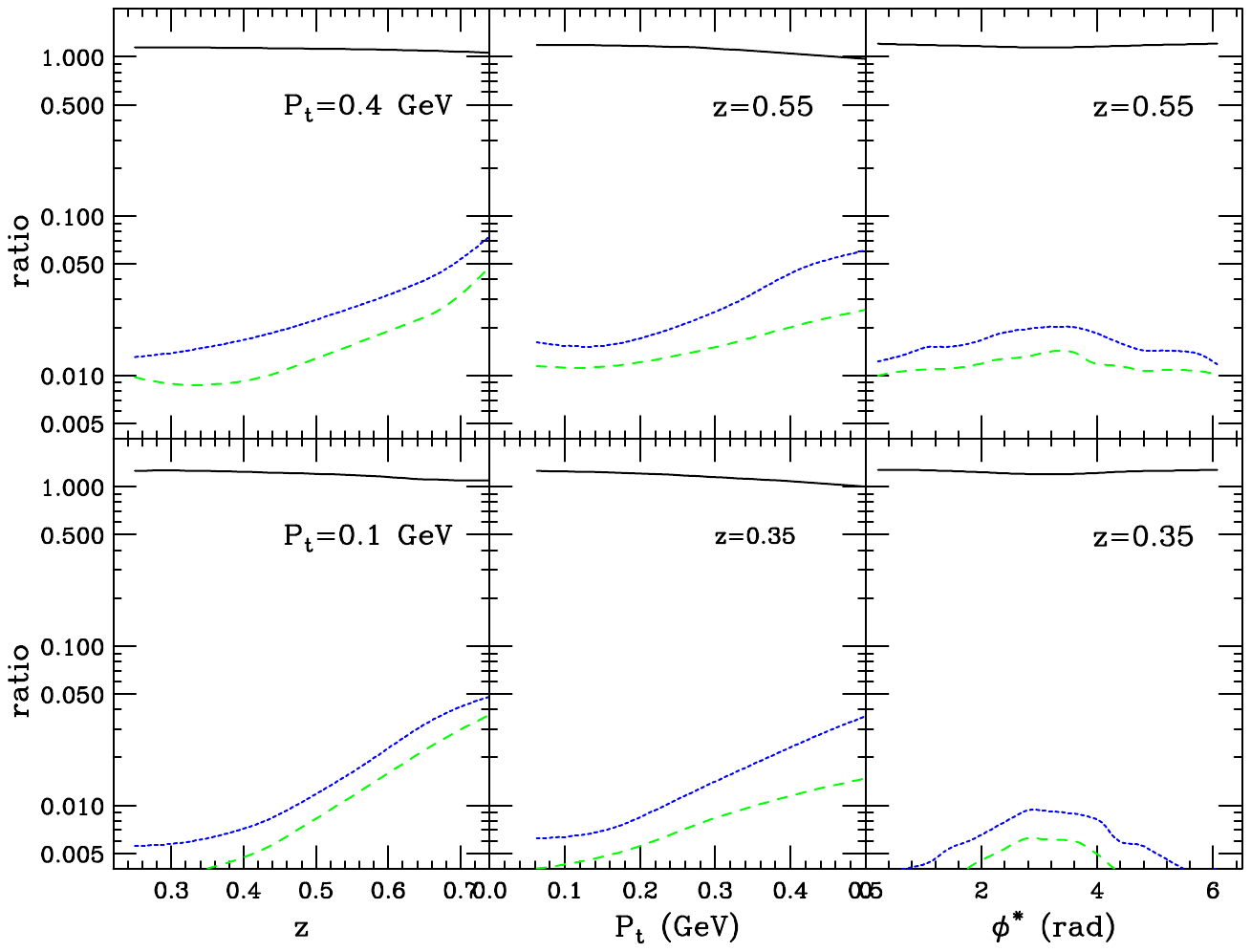}
 \caption{The solid black curves illustrate the ratio of radiated cross sections
 to Born cross sections 
 for $\pi^+$ from a deuteron target with  
 $x=0.3$ and $Q^2=3$ GeV$^2$. They are plotted in the left-hand panels 
 as a function of $z$ at $\phi^{*}=180^\circ$ for two values of $P_t$; in the
 middle panels as a function of $P_t$ for two values of $z$; and in the right-hand
 panels as a function of $\phi^*$ for two values of $z$.
 The short-dashed blue curves show the relative 
 contribution of the radiative tail from exclusive pion production, 
 while the long-dashed green curves show the contributions from the 
 $\pi\Delta$ radiative tail.} 
 
 \label{fig:rcall}
 \end{figure}


The $P_t$ and $z$ dependence of the radiative corrections follows a similar pattern, 
with the overall ratio decreasing at high $P_t$ or $z$, resulting from a 
strong increase of the exclusive and $\pi\Delta$ contributions being more than 
offset by a depletion of SIDIS events, as illustrated in the middle panels of Fig.~\ref{fig:rcall}.
The $\phi^{*}$ dependence of the radiative corrections indicates a small but 
non-negligible $\cos(\phi^{*})$ dependence, with the exclusive pion and $\pi\Delta$ 
contributions peaking near  180$^\circ$, as shown in the right-hand
panels of Fig.~\ref{fig:rcall}.

The exclusive pion and $\pi\Delta$ corrections were assumed to be subtractive, and reduced the measured cross section by as much as 7\% in the worst case. The multiplicative correction factor ranged from 0.8 at high $z$ to 1.2 at low $z$.

\subsection{ Acceptance corrections}
\begin{figure*}[htb!]
\includegraphics[width=0.75\textwidth]{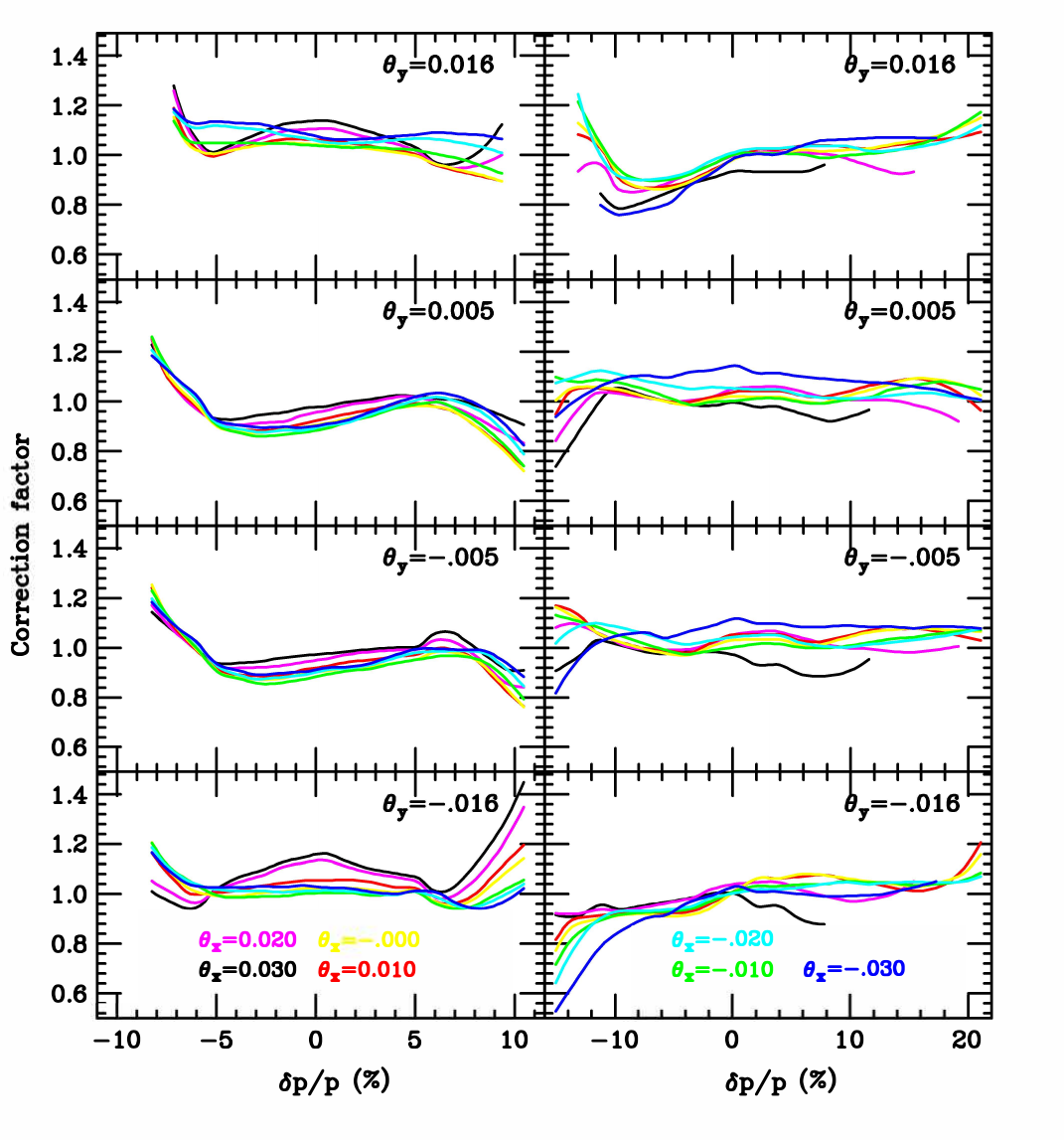}
\caption{Acceptance correction factors as a function of $\frac{p-p_0}{p_0}$ for four bins in in-plane relative scattering angle $\theta_y$ for 
HMS (left-hand column)  
and SHMS (right-hand column). The different colors correspond to
bins in out-of-plane angle $\theta_x$.}
\label{fig:acccorr}
\end{figure*}

 The predicted yields were corrected for small mismatches between the Monte Carlo simulation of the spectrometers and the actual acceptance. Three-dimensional grids in relative momentum ($\frac{p-p_0}{p_0}$), where $p_0$ is the central momentum, and the Euler
angles in the $x$  and $y$ directions ($x^{'}_{tar}$--vertical, and $y^{'}_{tar}$--pointing left) reconstructed to the frame of the target,
 were constructed by minimizing the $\chi^2$ to achieve agreement among the data taken at different central momenta and scattering angles, using the entire dataset of this experiment. The multiplicative correction factors, shown in Fig.~\ref{fig:acccorr}, are applied to the event-by-event weights for events generated in SIMC. For the HMS spectrometer, one of the prominent features is a ``dip" near $\frac{p-p_0}{p_0}=-2\%$ for the central scattering angles, with only minor dependence on out-of-plane angle. This feature has been noted before in previous one-dimensional studies that only looked at the dependence on $\frac{p-p_0}{p_0}$. This new 3D study shows that the ``dip" becomes more of a ``bump" at larger absolute values of $y^{'}_{tar}$, and also shows some non-trivial $x^{'}_{tar}$ dependence. The SHMS spectrometer was new for this experiment, so our acceptance study is the first one. We found little dependence on vertical angle in the region $-0.03<x^{'}_{tar}<0.03$ rad, where the bulk of the data reside. We found a considerable $\frac{p-p_0}{p_0}$ dependence which itself is significantly dependent on scattering angle. 

The SHMS and HMS momentum range used for this analysis was  -16\% $ < dp/p <$ +18\% and -9\% $< dp/p <$  +11\%, respectively, for central angles. For extreme values of in-plane and out-of-plane angles, for which acceptance corrections were more than $\approx$ 10\% away from unity, more restrictive $dp/p$ ranges were used. 
The SHMS and HMS in-plane relative angle range for this analysis was -30~mr~$< y^{'}_{tar} <$~30~mr, while the out of plane relative angle range was -55~mr~$<x^{'}_{tar}<$~55~mr and -65~mr$<x^{'}_{tar}<$~65~mr respectively. 

\subsection{Event selection cuts and efficiency corrections}
The SIMC weights were also corrected event-by-event for the detector efficiencies, which can vary with position in the spectrometer hut, especially for the heavy gas Cherenkov detector in the SHMS. The same event selection cuts were used on the SIMC track
positions at the HMS and SHMS detectors, spectrometer exit apertures, and reconstructed
momenta and angles as for the actual experimental data. An overall factor of 0.99 was
applied to account for pion absorption in the target. 

\begin{figure}[hbt!]
\centering
\includegraphics[width=0.75\textwidth] {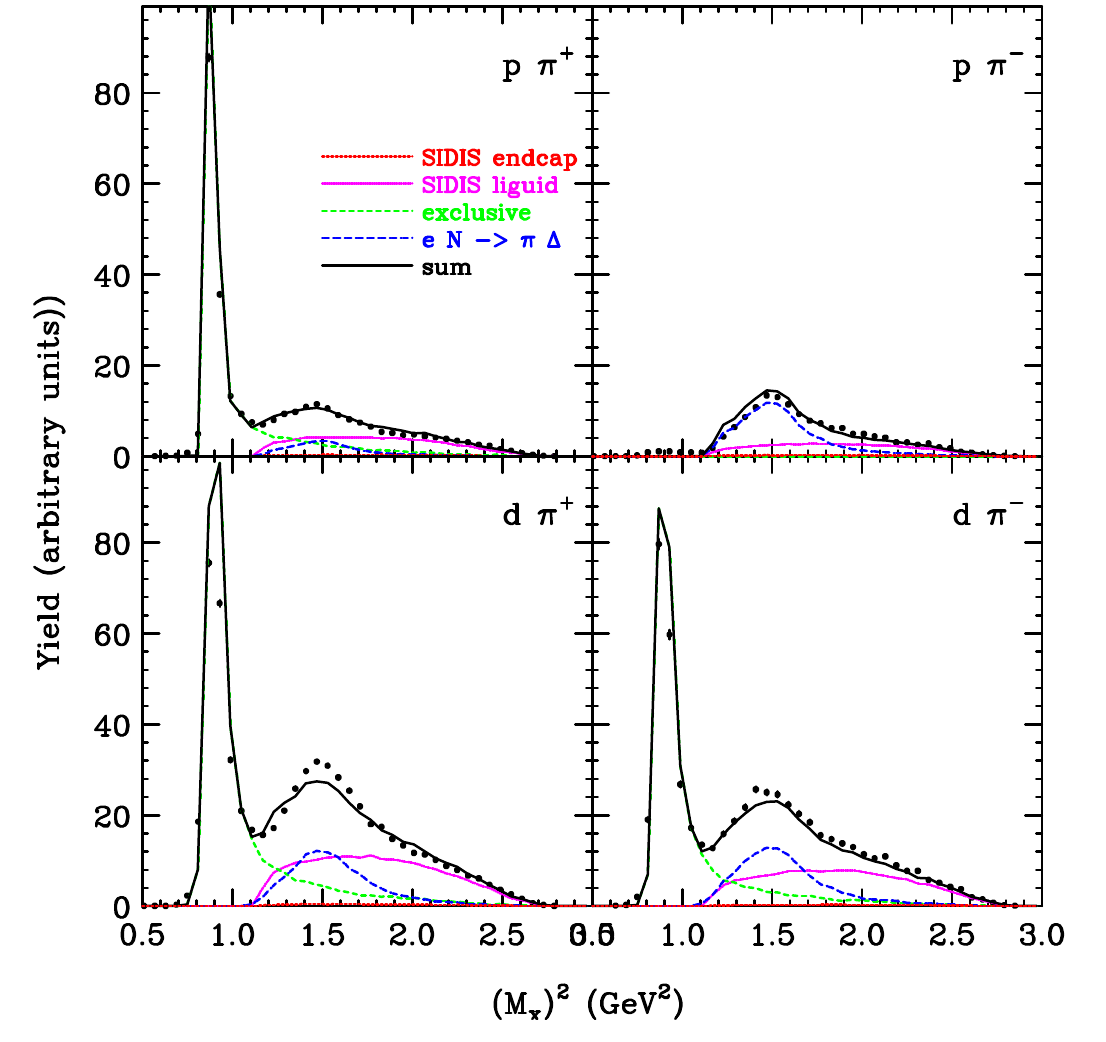}
 \caption{Experimental yields as a function of electron-pion missing
 mass squared for setting I from auxiliary runs taken with the SHMS
 centered on $z=0.9$, compared to the predicted Monte Carlo yields
 for SIDIS, exclusive pion production, and $\pi \Delta$ production. Results
 are shown for both $\pi^+$ (left panels) and $\pi^-$ production
 (right panels) and proton (top row) and deuteron targets (bottom row). The statistical uncertainty of the data points are smaller than the size of the points.}
 \label{fig:mmpi}
 \end{figure}

The quality of the models and corrections used in the simulation is demonstrated in Fig.~\ref{fig:mmpi}, showing the good agreement between the experimental yields and simulated yields
for setting I with the SHMS spectrometer centered on $z=0.9$ to
capture the contributions for exclusive pion production (centered on
electron-pion missing mass $M_x=0.94$ GeV), $\pi \Delta$ 
electroproduction (centered on $M_x=1.232$ GeV), and high-$z$ SIDIS
from both the target liquid and endcap (``dummy target'').

\section{Results}

\subsection{Data analysis}

 For each set of data with identical settings and target, 
 the number of electron-pion coincidences, corrected for
 accidental and target endcap contributions, were divided by the 
 accumulated beam charge to form an 
 experimental yield $Y_{exp}$. 
 
  The corrected yields were binned in a three-dimensional grid with twenty bins in
 $z$ from 0 to 1, 16 bins in $P_t$ from 0 to 1, and 15 bins in $\phi^{*}$
 from 0 to $2\pi$. 
 The predicted yields, $Y_{MC}$, from the MC simulation of each dataset were accumulated into the
 same kinematic grid as the experimental data. The simulated yields included contributions
 from SIDIS itself as well as the radiative tails from the 
 exclusive pion and $\pi \Delta$ reactions. The predicted yields were
 corrected for all detector and PID efficiencies as well as the luminosity
 dependence. The same detector position, magnet aperture, and reconstructed track 
 variables were used as for the experimental data. 

Experimental multiplicities, defined as the ratio of the SIDIS cross section ($d\sigma_{ee'\pi X}$) to the inclusive DIS cross section ($d\sigma_{ee'X}$), were determined for
each kinematic bin by:
\begin{equation}
M_i(z,P_t,\phi^{*})  = 
M_0(x, Q^2, z,P_t,\phi^{*}) \frac{Y_{exp}}{ Y_{MC}}
\end{equation}
for each target nucleus (p/d), HMS polarity, and $(x,Q^2)$ HMS 
setting, where $M_0$ is 
the multiplicity model used in the MC simulation, evaluated at
the center of each bin, and the index $i$ covers the SHMS settings
that provide overlap in $(z,P_t,\phi^{*})$. In most cases, there were
two overlapping settings, but occasionally there were three or four overlaps.
The final results were taken as the weighted average of all $M_i$ that contribute to each bin.

 The results discussed in this paper included the additional cut
 $M_x>1.6$ GeV, to remove the region where contributions from
 nucleon resonances, semi-exclusive processes, and higher-twist
 effects appear to be large, as was found in Ref.~\cite{Tigran07}.
 This cut was removed in the version of the analysis used to
 iterate the SIDIS model used in the Monte Carlo simulation.

 Numerical results for the multiplicities are tabulated in a full
 three-dimensional grid in $(z,P_t,\phi^*)$ for each
 target, pion polarity, and HMS setting in $(x,Q^2)$ on the
 Hall C experimental results web page~\cite{webpage}. In this table,
 each HMS spectrometer setting was divided in two, with relative
 scattering angle either positive or negative. A total of 20,000 bins
 are listed, based on the criterion that the Monte Carlo simulation
 prediction was for more than 4 counts, to ensure approximately
 Gaussian statistical errors on the experimental data. The table
 also includes results from thirteen additional HMS settings taken
 in Fall 2018 and Spring 2019 to study charge-symmetry violation in pion fragmentation functions, as
 reported in Ref.~\cite{Bhatt:2024prq}. These settings covered  a small 
 range $\left<P_t\right>\approx$ 0.1 GeV, and therefore are not included in the results of
 the present publication. The tables also include multiplicity results
 with no radiative corrections applied, which may prove useful
 in future global fits with consistent radiative correction models
 and formalism.


\subsection{Pion multiplicities as a function of $(z,P_t,\phi^*)$}

\begin{figure}[hbt!]
\centering
\includegraphics[width=1.0\textwidth,trim={0cm 0cm 0.5cm 0cm},clip] {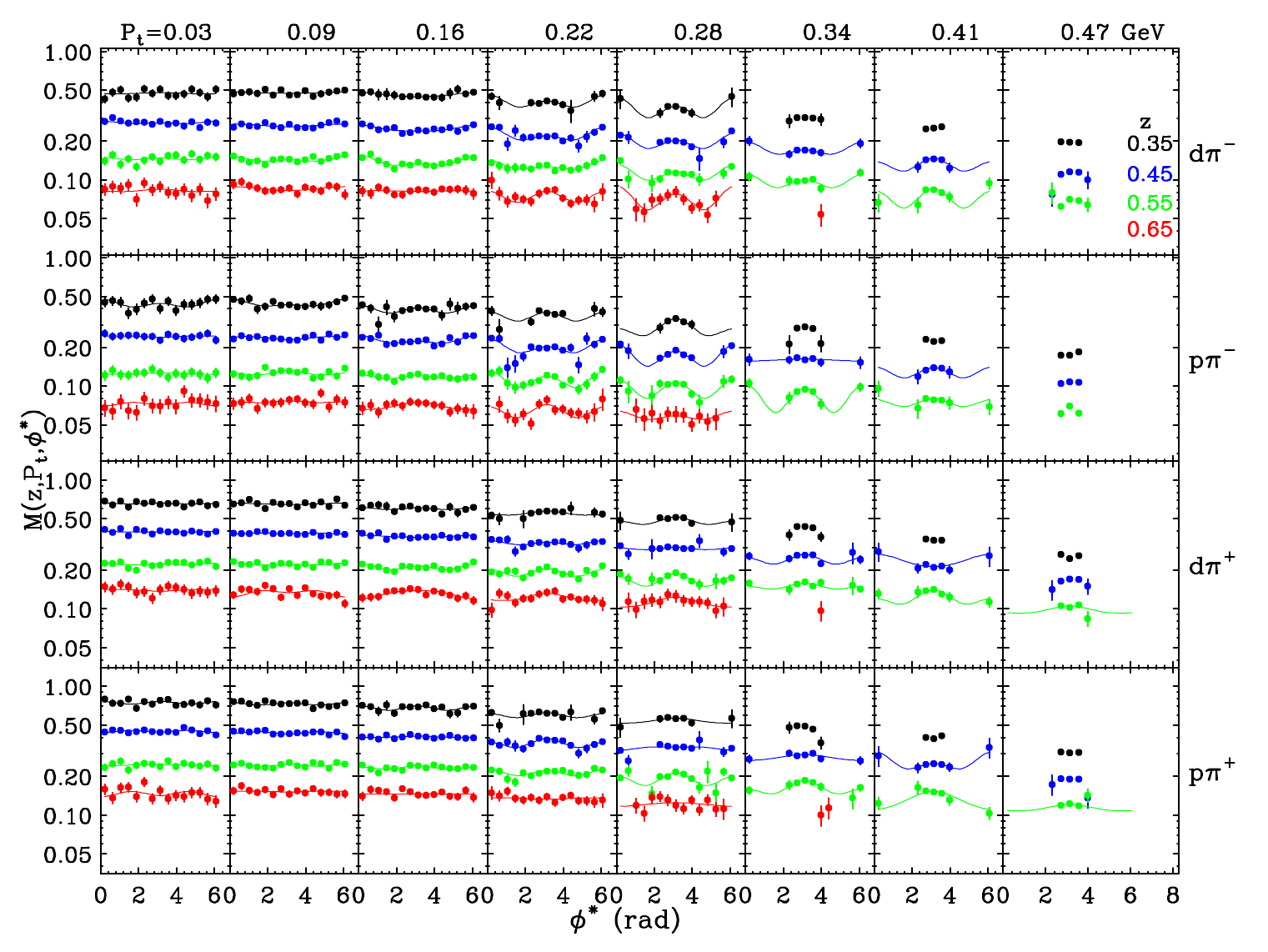}
 \caption{Pion multiplicity as a function of $\phi^{*}$ for $x = 0.31$, $Q^2=3$ GeV$^2$ (kinematic setting I) for eight bins in $P_t$ (left to right) and four target/final-state configurations (top to bottom), for four values of $z$ as indicated on the right edge of the rightmost panels. Relative systematic errors (not shown) are estimated to
 be 2.2\% point-to-point
 and 1.8\% for the scale uncertainty. The solid curves are fits to each dataset at fixed $z$, $P_t$, target, and pion charge with the functional form 
  $M_0  [1 + A \cos(\phi^{*}) + B\cos(2\phi^{*})]$. No fully differential results
  from previously published experiments are available to compare to.}
 \label{fig:mpt1}
 \end{figure}
 
 \begin{figure}[hbt!]
\centering
\includegraphics[width=1.0\textwidth,trim={0cm 0cm 0cm 0.0cm},clip] {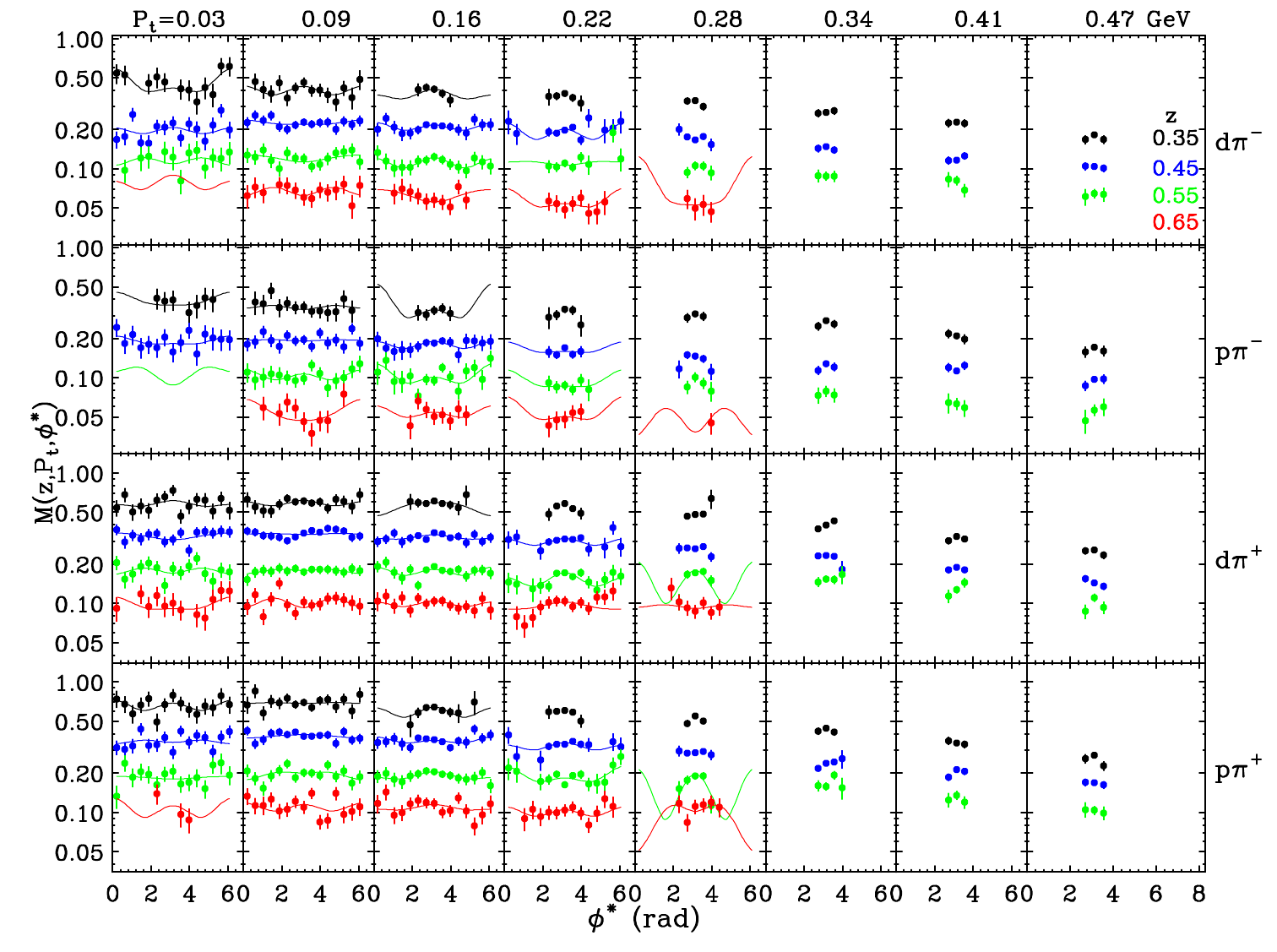}
 \caption{Pion multiplicity as a function of $\phi^{*}$ for $x = 0.30$, $Q^2=4.1$ GeV$^2$ (kinematic setting II) for eight bins in $P_t$ (left to right) and four target/final-state configurations (top to bottom), for four values of $z$ as indicated on the right edge of the rightmost panels. Relative systematic errors (not shown) are estimated to
 2.2\% point-to-point 
 and 1.8\% for the scale uncertainty. The solid curves are fits to each dataset at fixed $z$, $P_t$, target, and pion charge with the functional form 
  $M_0  [1 + A \cos(\phi^{*}) + B\cos(2\phi^{*})]$. No fully differential results
  from previously published experiments are available to compare to.}
  \label{fig:mpt2}
 \end{figure}
 
\begin{figure}[hbt!]
\centering
\includegraphics[width=1.0\textwidth,trim={0cm 0cm 0cm 0.0cm},clip] {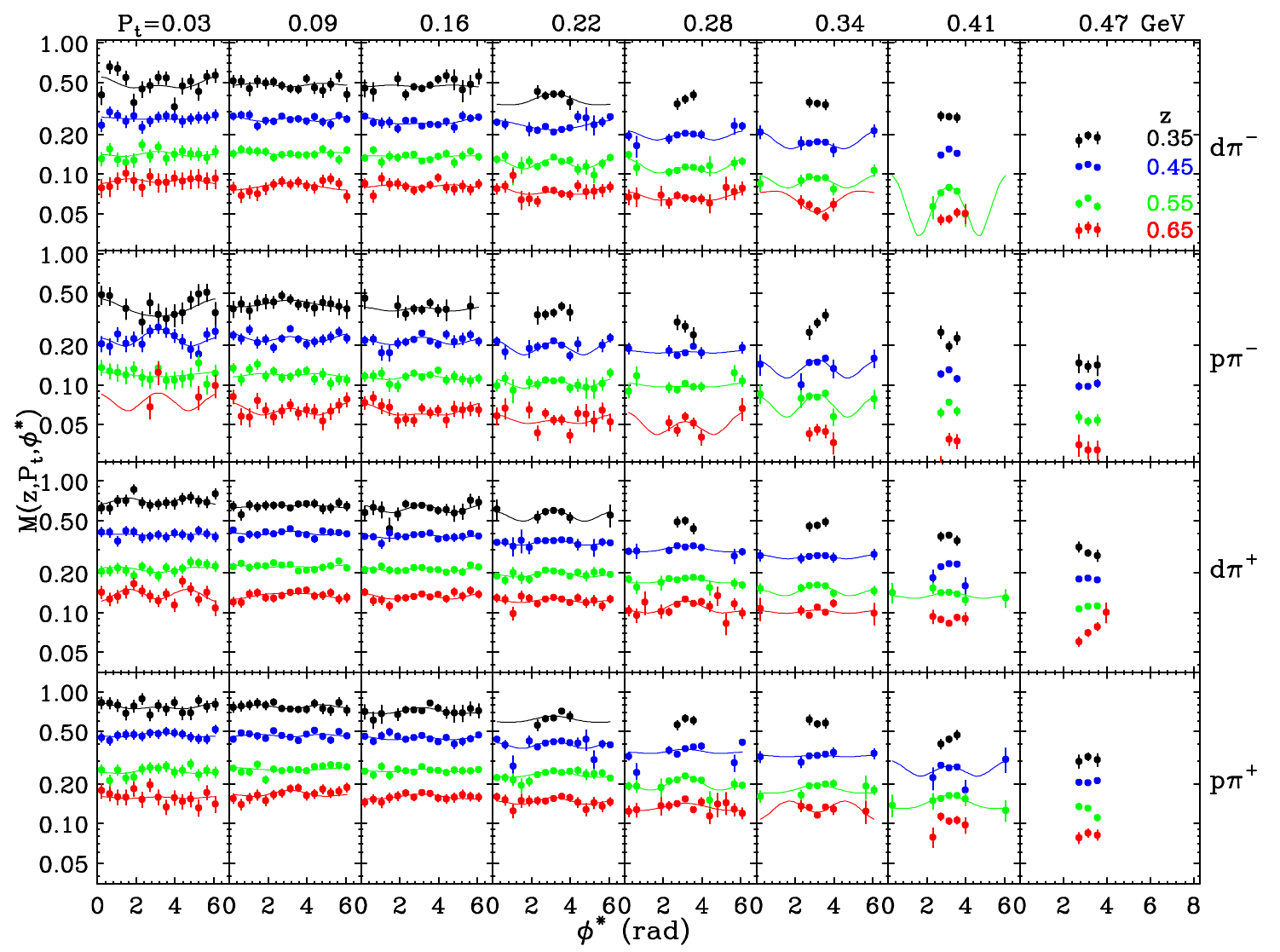}
 \caption{Pion multiplicity as a function of $\phi^{*}$ for $x = 0.45$, $Q^2=4.5$ GeV$^2$ (kinematic setting III) for eight bins in $P_t$ (left to right) and four target/final-state configurations (top to bottom), for four values of $z$ as indicated on the right edge of the rightmost panels. Relative systematic errors (not shown) are estimated to
 be 1.3\% point-to-point
 and 1.8\% for the scale uncertainty. The solid curves are fits to each dataset at fixed $z$, $P_t$, target, and pion charge with the functional form 
  $M_0  [1 + A \cos(\phi^{*}) + B\cos(2\phi^{*})]$. No fully differential results
  from previously published experiments are available to compare to.}
  \label{fig:mpt3}
 \end{figure}

The $\phi^{*}$ dependence of the semi-inclusive pion electroproduction multiplicity $M(x,Q^2,z,P_{t},\phi^{*})$ is shown in discrete bins of $z$ and $P_t$ for kinematic setting I 
($x=0.31$, $Q^2=3.1$ GeV$^2$, $W=2.8$ GeV) in Fig.~\ref{fig:mpt1}. For clarity, adjacent bins in $z$ were combined together, 
and only eight bins in $P_t$ are shown; at higher values of $P_t$, the $\phi^{*}$ coverage becomes increasingly centered near 180 degrees due to the use of in-plane spectrometers in this experiment.
Similarly, the results for settings II and III are shown in Fig.~\ref{fig:mpt2} and
Fig.~\ref{fig:mpt3} respectively.

The main features of the data are: \\a) the multiplicity decreases
with increasing $z$;\\ b) the multiplicity decreases with increasing 
$P_t$; and \\c) the distributions tend to be mostly independent of 
$\phi^{*}$ at fixed values of $z$ and $P_t$. 

To quantify this behavior, each dataset at fixed 
$z$, $P_t$, target, and pion charge was fit with the functional 
form 
\begin{equation}
M_0  [1 + A \cos(\phi^{*}) + B\cos(2\phi^{*})].
\label{eq:phidep}
\end{equation}

In terms of the standard structure functions~\cite{Bacchetta07}, 
$$M_0=(F_{UU,T} +\epsilon F_{UU,L}) /(F_{T} +\epsilon F_{L})$$
$$A = \sqrt{2\epsilon(1+P_t)}F^{\cos(\phi^{*})}_{UU}/(F_{UU,T} +\epsilon F_{UU,L})$$
$$B = P_t F^{\cos(2\phi^{*})}_{UU}/ (F_{UU,T} +\epsilon F_{UU,L})$$

 The fit results are discussed in the next subsections.




\subsection{Pion multiplicities averaged over $\phi^{*}$}
\label{sec:mult_phiav}
\begin{figure}[htb!]
\centering
\includegraphics[width=0.9\textwidth,trim={0cm 0cm 0cm 0cm},clip] {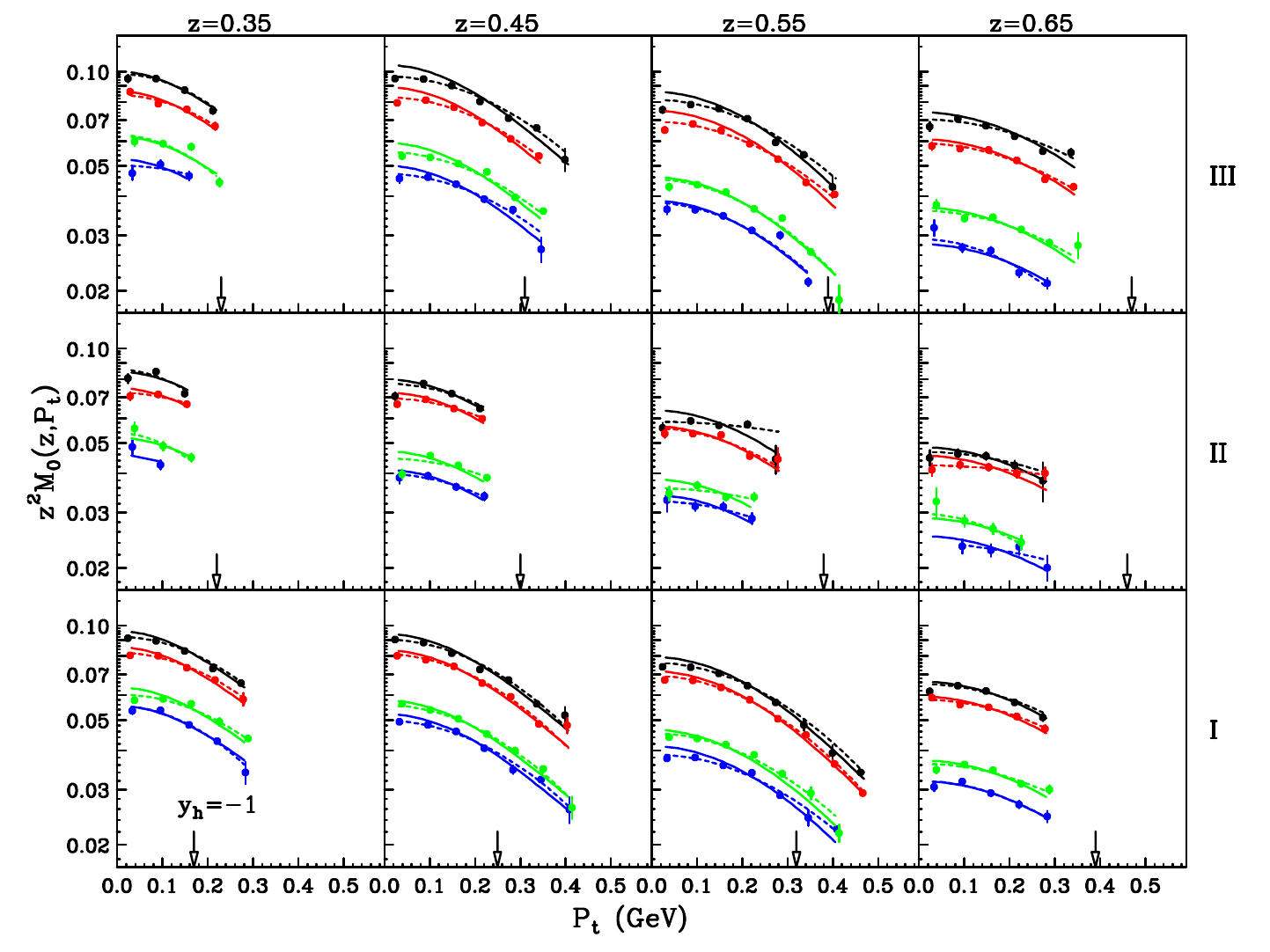}
 \caption{The $\phi^*$ averaged fit parameter $M_0$ weighted by $z^2$, as a function of $P_t$, for the three kinematic settings of this experiment (top to bottom) in four bins in $z$ (left to right). Within each panel, the results from top to bottom are for $\pi^+$ from a proton target (black), $\pi^+$ from a deuteron target (red), $\pi^-$ from a deuteron target (green) and $\pi^-$ from a proton target (blue). Relative systematic errors (not shown) are estimated to
 be 1.3\% (2.2\%) point-to-point for setting III (settings I and II) 
 and 1.8\% for the scale uncertainty. The curves are the predictions of the MAP collaboration based on a fit to previous world data \cite{MAPS,MAP24,MAP22_4}, normalized with a parameter $k$ to give the best overall agreement with these results. The dashed curves illustrate two-parameter Gaussian fits to each dataset of the form $M=a e^{(-b P_t^2)}$. The arrows point to the values of $P_t$ for which $y_h=-1$, where $y_h$ is the definition of rapidity used in Ref. \cite{Boglione:2022gpv}.}
 \label{fig:ptmm0w}
 \end{figure}
The results for the $\phi^{*}$ averaged parameter, $M_0$, from the fits described above are displayed in Fig.~\ref{fig:ptmm0w} as a function
of $P_t$ for the three kinematic settings, the target and pion charge combinations, in
four bins in $z$. The measured multiplicities are compared to the calculation of MAP~\cite{MAPS,MAP24,MAP22_4} scaled by $P_t$-independent  normalization factors $k$ that give the best agreement with these data. 
The MAP calculations, which use a combination of Gaussian and weighted Gaussian distributions in $P_t$, based on a fit to data from HERMES and COMPASS, are generally in good agreement with the measured $P_t$-dependence.  
\begin{figure}[htb!]
\centering
\includegraphics[width=0.55\textwidth] {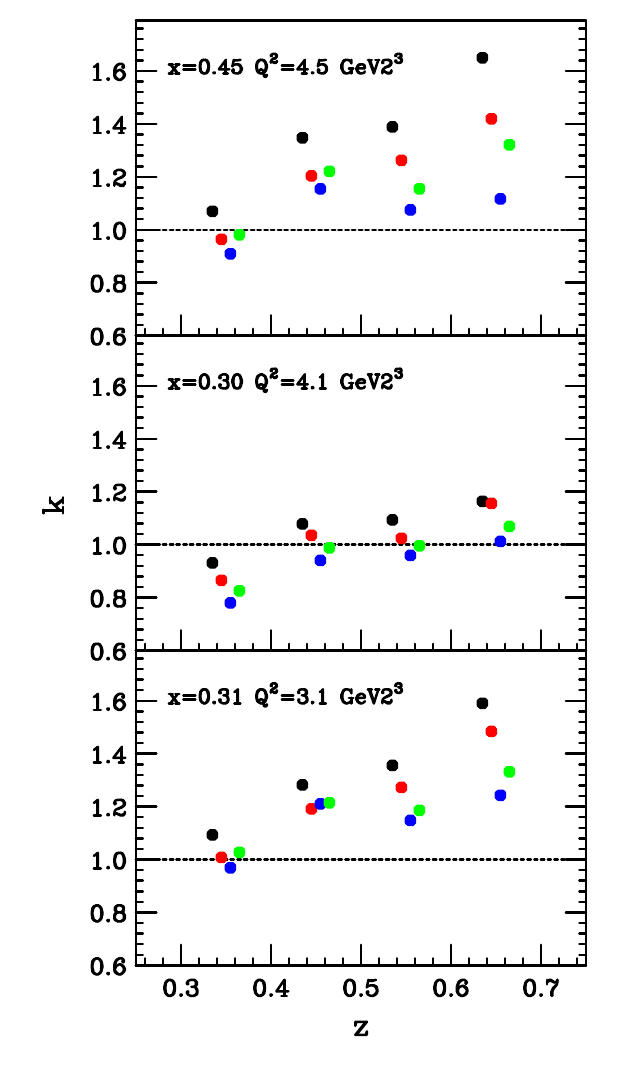}
 \caption{Normalization factors $k$ that best match our results and the MAP calculations~\cite{MAPS,MAP24,MAP22_4}
 for the three kinematic settings of this experiment (top to bottom), as a function of $z$,  and the four flavor cases (color scheme same as in Fig.~\ref{fig:ptmm0w}).}
 \label{fig:kf}
 \end{figure}
 Also shown in
Fig.~\ref{fig:ptmm0w} are Gaussian fits of the form $a e^{-b P_t^2}$. It can be seen that the fits are roughly parallel to each other when plotted on a logarithmic 
scale, indicating only small differences in the slope parameters $b$. In the simple
ansatz that $b^{-1}= { \langle\vec{p}^{2}_{\perp}\rangle + z^{2} \langle\vec{k}^{2}_{T}\rangle}$, this implies small differences in $\langle\vec{k}^{2}_{T}\rangle$
for  up and down quarks, as well as small difference in the widths of favored fragmentation and unfavored fragmentation functions. 
This is discussed further in the next section. 

The $P_t$-independent normalization factors $k$ are plotted in Fig.~\ref{fig:kf} as a function of $z$,  for the target and pion charge combinations and the three kinematic settings. They are, on average, closest to unity for setting II ($W = 3.3$ GeV), and tend to be larger than unity for setting I ($W = 2.8$ GeV), and even larger for setting III ($W = 2.6$ GeV). There is also a clear trend for $k$ to increase with increasing $z$, especially for positive pions from the proton target, and to a lesser extent for the positive pions from the deuteron 
target.  These trends  are likely related to the fact that the present data are at lower $W$ and higher $x$ than the HERMES and COMPASS data that went into the MAP global fit. 

\subsection{Pion multiplicities near $\phi^{*}=180^\circ$}
\begin{figure}[htb!]
\centering
\includegraphics[width=1.0\textwidth] {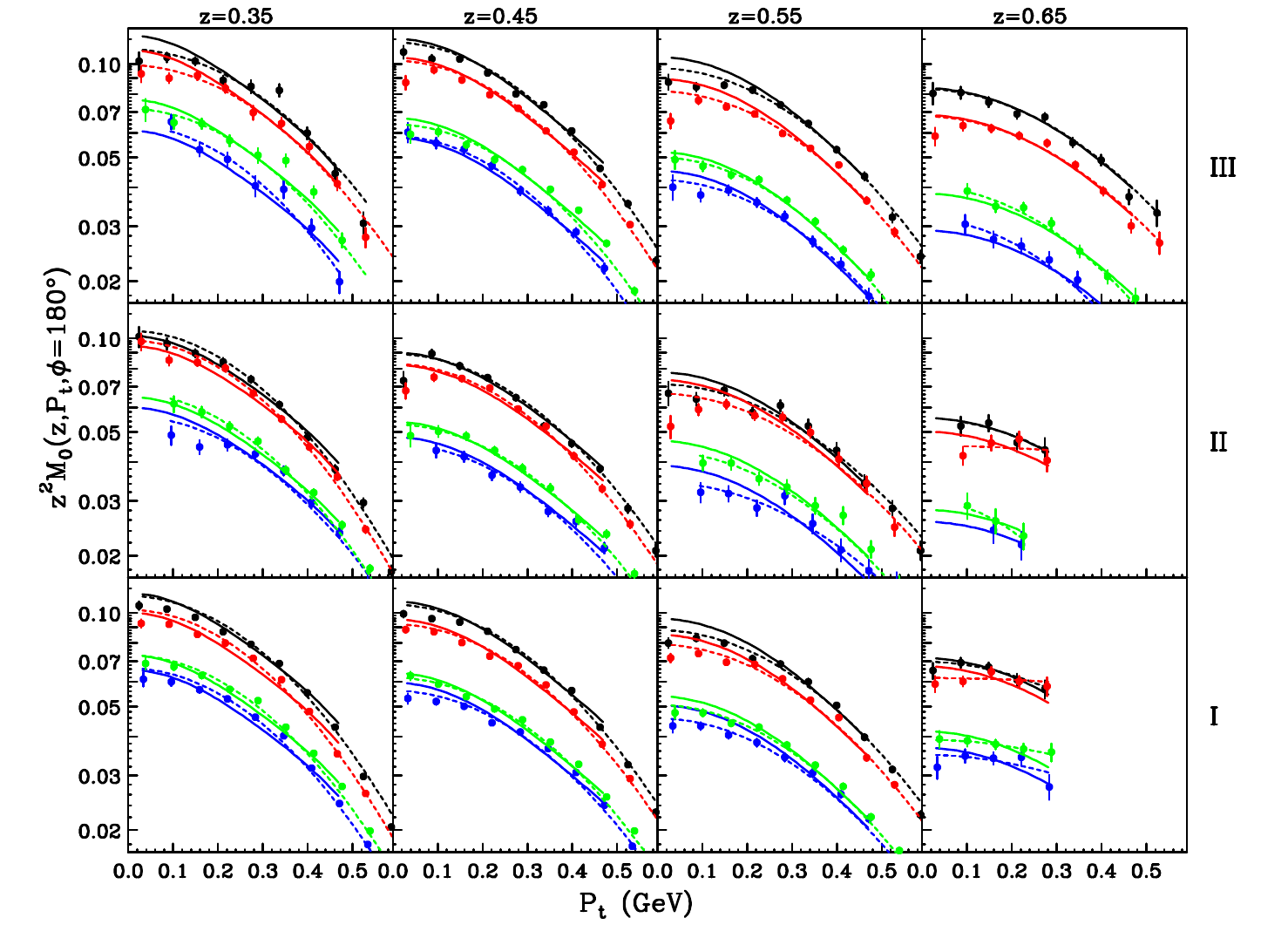}
 \caption{The fit parameter $M_0$ weighted by $z^2$ as a function of $P_t$, averaged over $150<\phi^*<210$ degrees, for the three kinematic settings of this experiment (top to bottom) in four bins in $z$ (left to right). Within each panel, the results from top to bottom are for $\pi^+$ from a proton target (black), $\pi^+$ from a deuteron target (red), $\pi^-$ from a deuteron target (green) and $\pi^-$ from a proton target (blue). 
Relative systematic errors (not shown) are estimated to
 be 1.3\% (2.2\%) point-to-point for setting III (settings I and II) 
 and 1.8\% for the scale uncertainty. 
 The curves are the predictions of the MAP collaboration based on a fit to previous world data \cite{MAPS,MAP24,MAP22_4}, normalized with a parameter $k$ to give the best overall agreement with these results. The dashed curves illustrate two-parameter Gaussian fits to each dataset of the form $M=a e^{(-b P_t^2)}$. No fully differential results from previously published experiments are available for comparison.} 
 \label{fig:ptm180w}
 \end{figure}
 Due to the experimental setup limitations, it was not possible to obtain full azimuthal coverage at large $P_t$. Nonetheless, a considerable amount of time was spent accumulating data 
 near $\phi^{*}=$180$^\circ$, which primarily measures $M_0 (1 - A + B)$. Given this caveat, it is of interest to plot the multiplicities at $\left<\phi^{*}\right> \approx $ 180$^\circ$, over the full $P_t$ experimental range, as shown in 
  Fig.~\ref{fig:ptm180w}. The MAP calculations for $M_0$ (MAP did not include non-zero values of A and B in their fits) describe the $P_t$-independence quite well
  up to $P_t=0.5$ GeV, which is the largest value provided by the collaboration. Although not perfect in magnitude, the Gaussian-shaped fits shown in the figure describe the
  $P_t$-independence remarkably well to values of $P_t$ as high as 0.7 GeV.
\begin{figure}[htb!]
\centering
\includegraphics[width=0.45\textwidth] {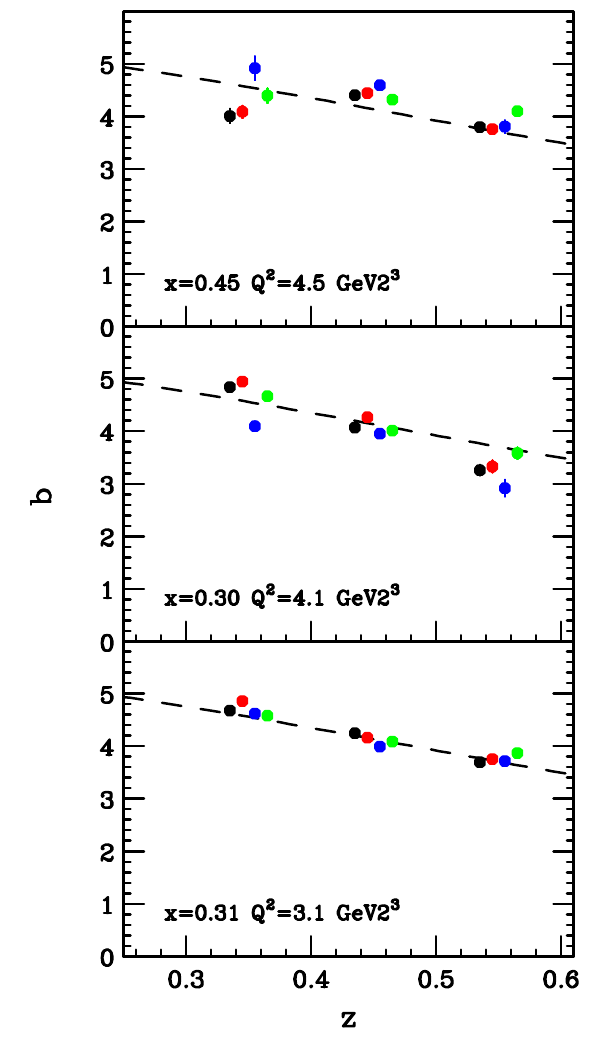}
 \caption{Slope parameters $b$ for the fits shown Fig.~\ref{fig:ptm180w} as
a function of $z$
 for the three kinematic settings of this experiment (top to bottom) and the four flavor cases (color scheme same as in Fig.~\ref{fig:ptmm0w}).}
 \label{fig:g2}
 \end{figure}

The slope parameters $b$ for the above fits are displayed in Fig.~\ref{fig:g2} as
a function of $z$. The values decrease with increasing $z$, with a rough global 
fit to all three kinematic settings given by $b=(0.185 + z^2 0.28)^{-1}$ GeV$^{-2}$.
With the simple ansatz $b^{-1}= { \langle\vec{p}^{2}_{\perp}\rangle + z^{2} \langle\vec{k}^{2}_{T}\rangle}$, the up and down quark widths $\langle\vec{k}^{2}_{T}\rangle \approx 0.28$ GeV$^{-2}$ and the fragmentation widths $\langle\vec{p}^{2}_{\perp}\rangle\approx 0.185$ GeV$^{-2}$. As mentioned in the previous section, this is consistent with little or no differences in $\langle\vec{k}^{2}_{T}\rangle$
for  up and down quarks, as well as the widths of favored and unfavored fragmentation functions. 

\begin{figure}[htb!]
\centering
\includegraphics[width=1.0\textwidth] {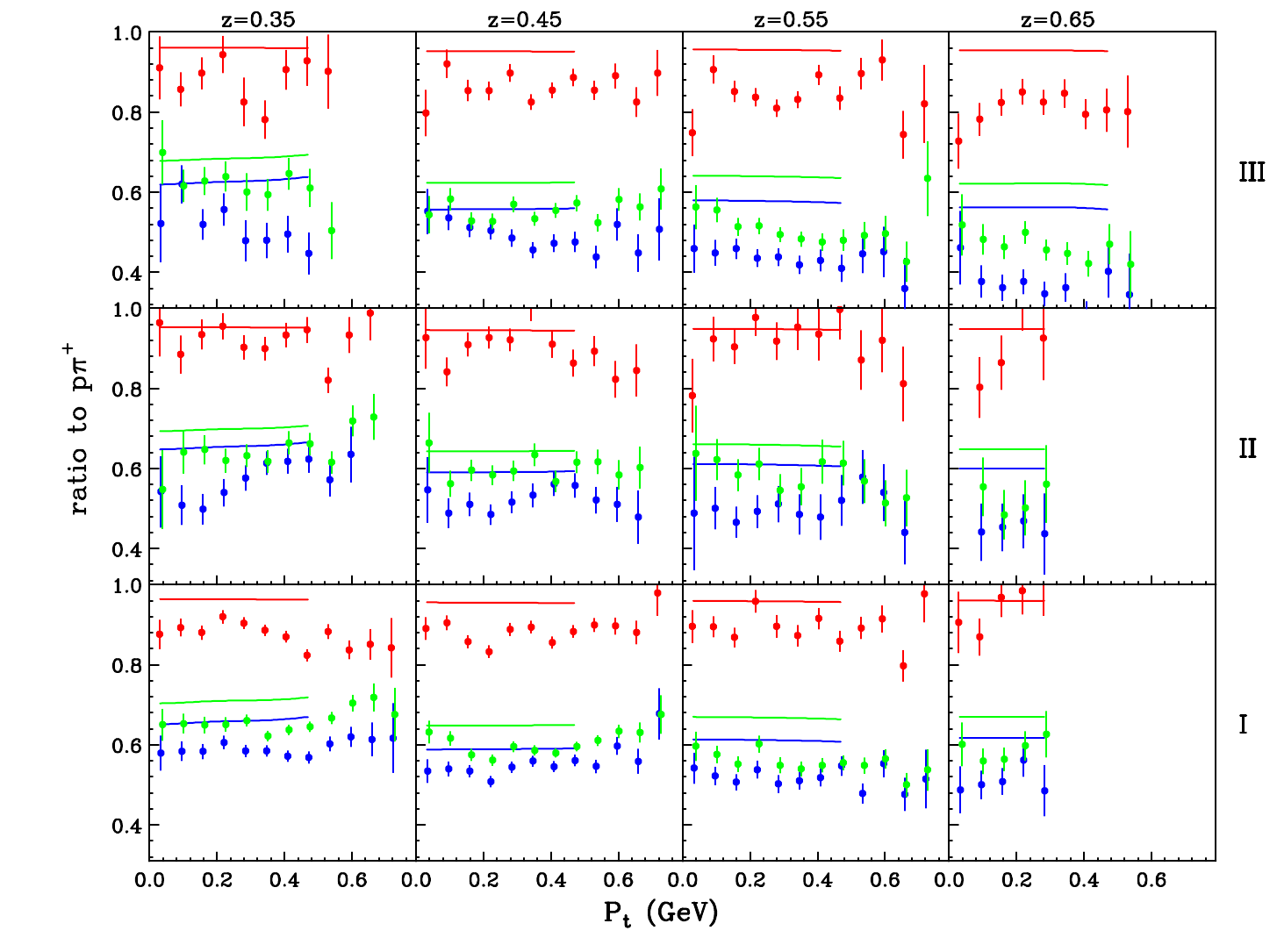}
 \caption{Ratios to $(p,\pi^+)$  of experimental multiplicities averaged over $150<\phi^{*}<210$ degrees for $(d,\pi^+)$ (red), $(p,\pi^-)$ (green), 
 and $(d,\pi^-)$ (blue). Systematic errors (not shown) are discussed in the text. 
 The solid curves are the ratios of MAP calculations, while the dashed curves are the ratios of Gaussian fits from Fig.~\ref{fig:ptm180w}. The curves follow the same color scheme as the data points.} 
 \label{fig:ptmrat180w}
 \end{figure}
 
 To further quantify the observed flavor independence of the multiplicities, the ratio of $\left<\phi^{*}\right> \approx $ 180$^\circ$ multiplicities for 
$(d,\pi^+)$, $(p,\pi^-)$, and $(d,\pi^-)$ to the multiplicities for  $(p,\pi^+)$ are plotted
in Fig.~\ref{fig:ptmrat180w}, as a function of $P_t$ for the three kinematic settings and 
four $z$ bins. The ratios are consistent with no statistically significant dependence on $P_t$. The ratios are
compared to those from the MAP calculations~\cite{MAPS,MAP24,MAP22_4}, which also show only
very slight $P_t$-dependence. 

\subsection{Azimuthal dependence of multiplicities}
\begin{figure}[hbt!]
\centering
\includegraphics[width=1.0\textwidth] {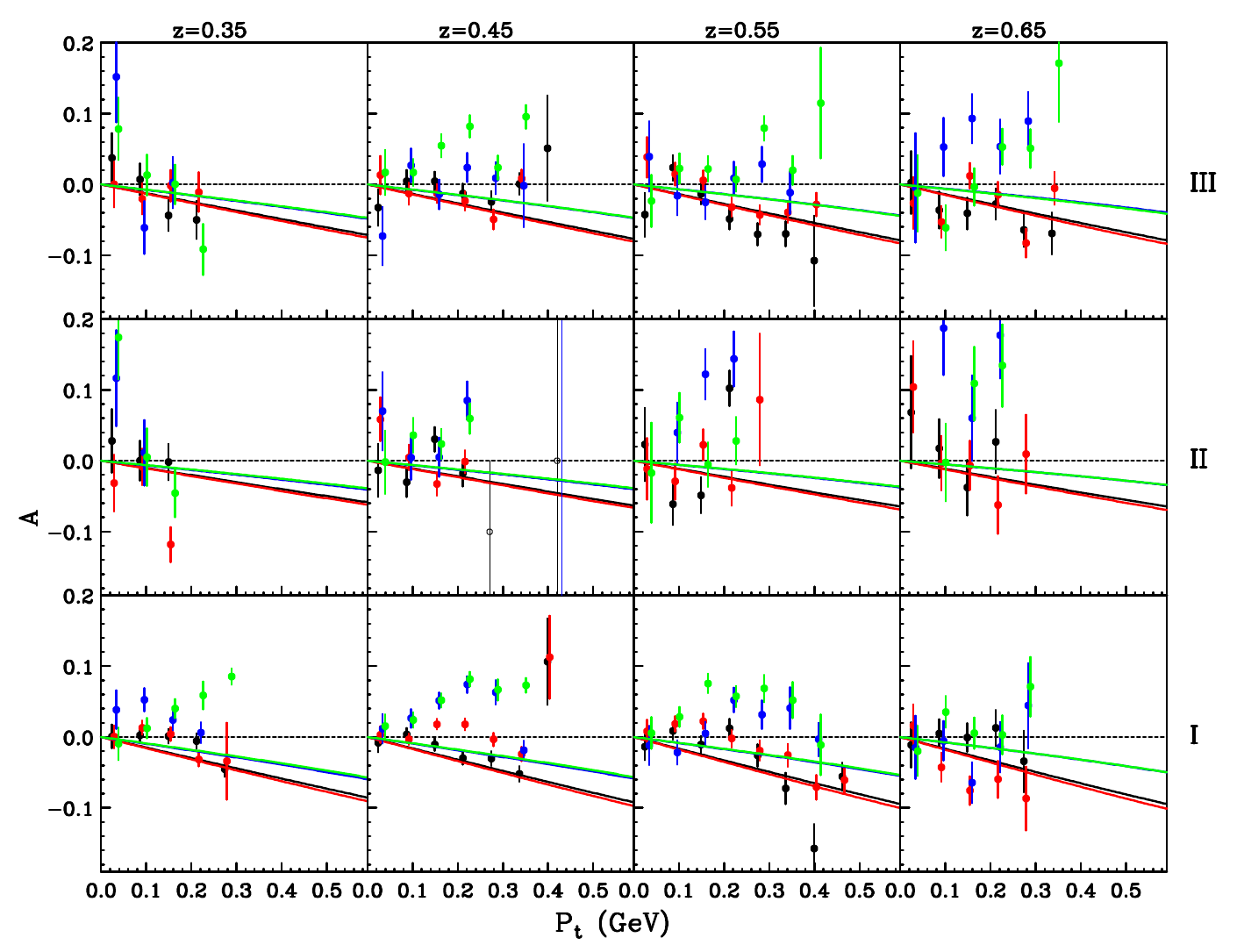}
 \caption{Results from this experiment (solid circles) 
 for the coefficients $A$ that scale the $\cos(\phi^{*})$ distributions for $(p,\pi^+)$ (black),  $(d,\pi^+)$ (red), $(p,\pi^-)$ (green), 
 and $(d,\pi^-)$ (blue). Relative systematic errors (not shown) are estimated to
 be 0.01 (0.02) point-to-point for setting III (settings I and II). The scale
 uncertainty is negligibly small. 
 The open circles are from HERMES at $Q^2=6.6$ GeV$^2 
 $, $x=$0.36 and $\left<z\right>=$0.45\cite{Barone:2015ksa}, and curves are from a fit to HERMES and COMPASS data~\cite{Barone:2015ksa, HERMES:2012kpt}. The curves follow the same color scheme as the data points.} 
  \label{fig:ptmphiw}
 \end{figure}

\begin{figure}[hbt!]
\centering
\includegraphics[width=1.0\textwidth] {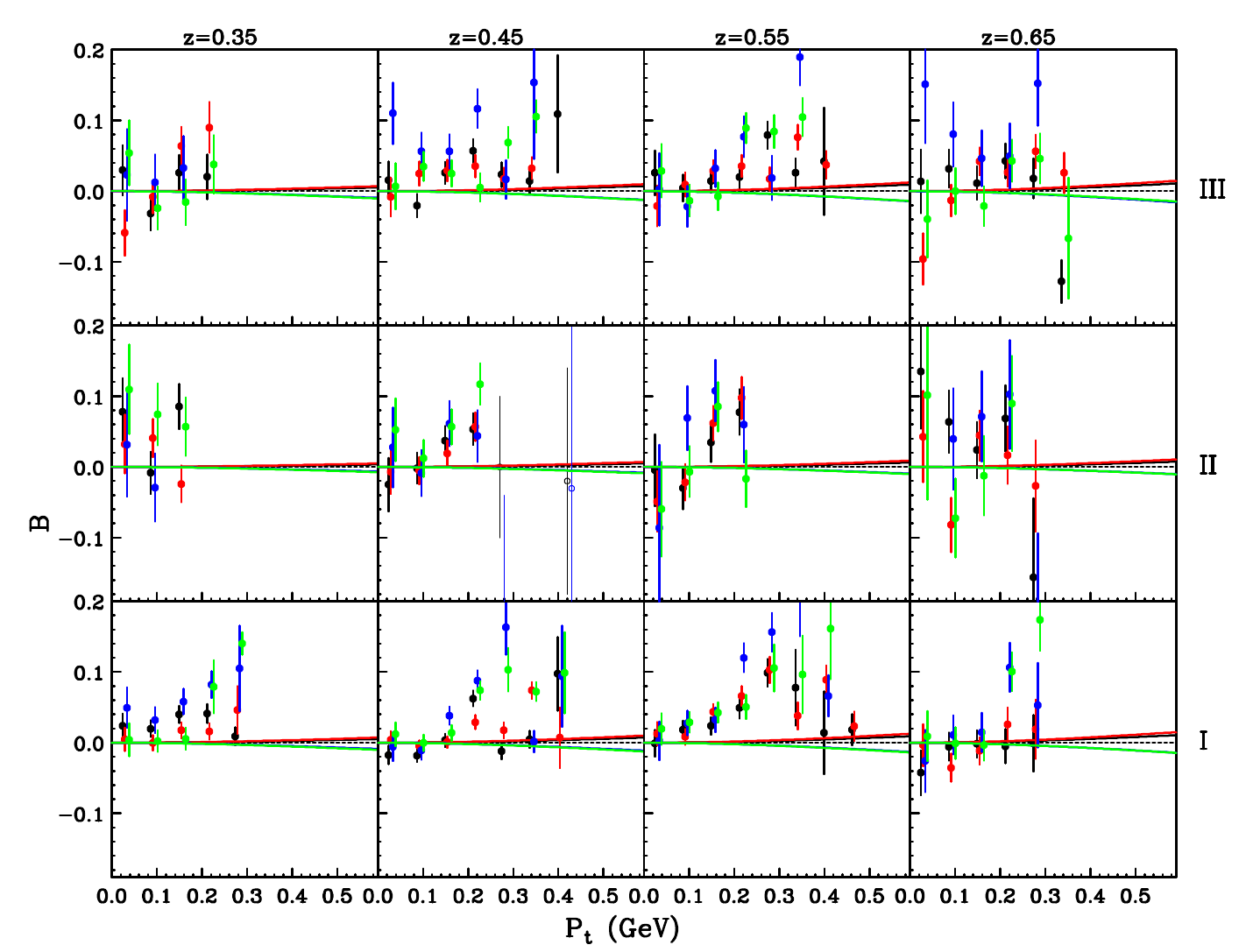}
 \caption{ Results from this experiment (solid circles) 
 for the coefficients $B$ that scale the $\cos(2\phi^{*})$ distributions for $(p,\pi^+)$ (black),  $(d,\pi^+)$ (red), $(p,\pi^-)$ (green), 
 and $(d,\pi^-)$ (blue). Relative systematic errors (not shown) are estimated to
 be 0.01 (0.02) point-to-point for setting III (settings I and II). The scale
 uncertainty is negligibly small. The open circles are from HERMES at $Q^2=6.6$ GeV$^2 
 $, $x$=0.36 and $\left<z\right>=$0.45\cite{Barone:2015ksa}, and curves are from a fit to HERMES and COMPASS data~\cite{Barone:2015ksa, HERMES:2012kpt}. The curves follow the same color scheme as the data points.} 
  \label{fig:ptm2phiw}
 \end{figure}

The $\phi^{*}$ dependence of the measured multiplicities is quantified by the two coefficients, $A$ and $B$, associated with the $\cos(\phi^{*})$ and $\cos{(2\phi^{*})}$ modulations of the multiplicities. The $\cos(\phi^{*})$ coefficient,  $A$, obtained from the fit of the multiplicity results in each $(z,P_t)$ bin  to functional form-- Eq.~\ref{eq:phidep}, is shown in  Fig.~\ref{fig:ptmphiw} as a function of $P_t$.
These results show an overall trend that $A$ for $\pi^-$ production on both protons and deuterons is significantly~$>$~0 at high $z$ for all three kinematic settings. On the other hand, the $A$ coefficient is consistent with zero within experimental uncertainties or has small negative values for $\pi^+$ production on both protons and deuterons. The $\pi^+$ results are consistent with the previous HERMES measurements ~\cite{HERMES:2012kpt} but have the opposite sign for the $\pi^-$, except for the HERMES point~\cite{HERMES:2012kpt} at highest $x$. We note that all of these data generally lie at higher $x$ than those from HERMES~\cite{HERMES:2012kpt} or COMPASS~\cite{COMPASS_PT_ref}.
Similarly the $\cos{(2\phi^{*})}$ coefficient,  $B$, obtained from the fit of the multiplicity results in each $(z,P_t)$ bin  to Eq.~\ref{eq:phidep} is shown in  Fig.~\ref{fig:ptm2phiw} as a function of $P_t$. Other than a couple of $z$ bins, these results show small values of $B$, that are either consistent with zero or have small positive values. These results are consistent with the previous HERMES measurements~\cite{HERMES:2012kpt} except at $\left<z\right> =$0.55.

\section{Systematic Uncertainties and caveats}

\subsection{Experimental systematic studies}

The entire analysis was performed with several alternative sets of cuts
and PID criteria. One study used a smaller range in track momenta 
and angles. Another ignored the heavy gas Cherenkov detector in the SHMS, which
resulted in significant kaon contamination of the pion sample above
momenta of 3 GeV. A third study used a lower aerogel threshold, with 
correspondingly larger kaon contamination subtraction. The luminosity and HMS acceptance were verified to be accurate to within
3\% by comparison of elastic electron-proton measurements to a fit
to global data~\cite{bosted_fit}. 
The optical properties of the spectrometers were verified using the 
kinematic constraints of both $ep$ elastic scattering and exclusive
pion production. 

\subsection{Experimental systematic errors}
The main sources of systematic uncertainties are 
listed in Table~\ref{tab:syst}, based on the studies mentioned above
as well as known instrumental uncertainties. They have been divided 
into two categories: normalization/scale uncertainties that impact all 
measurements on a given target equally, and point-to-point uncertainties that vary with
pion kinematics and charge. The overall experimental systematic error
is estimated to be about 2.5\% for setting III. Due to many
problems in Spring 2018, we estimate an additional overall normalization
error of 2\% for settings I and II. 

In ratios of $\pi^+/\pi^-$, most systematic errors cancel out except for
the uncertainty in particle identification.
In ratios of proton to deuteron 
target for a fixed flavor, most systematic errors cancel except for
the relative target thickness.

\begin{table}[!hbt]
    \caption{Principal experimental systematic uncertainties, divided
    in overall normalization (scale) uncertainties and those that vary
    with pion kinematics.}
    \label{tab:syst} 
    \centering
    \begin{tabular}{|l|c|c|}
    \hline
        Source & Scale  & Point-to-Point  \\
               & Uncertainty (\%) & Uncertainty (\%) \\
        \hline
        Charge &- & 0.5 \\
        Target density & 1 & - \\
        Target boiling correction &- & 0.3\\
        Target end cap subtraction & 0.3 &- \\
         Particle identification & 1 & - \\
        PID Purity & -& 0.2 \\
        Spectrometer Acceptance &1 & 0.5 \\
        Kinematics & -& 0.3 \\
        Rate dependence & - & 1-2 \\
 
        \hline
        Total & 1.8 & 1.3-2.2 \\
        \hline
    \end{tabular}
\end{table}

\subsection{Radiative corrections uncertainty}

The application of radiative corrections is ideally an iterative
process in which all available global data are iteratively analyzed until
convergence is achieved. For the present analysis, we rely 
on our fits to the world data (including our own) on three physics
processes: exclusive pion production, $\Delta$ resonance production, and $\rho$ meson production. Our fit to exclusive pion production is driven
largely by preliminary, unpublished results from Hall C experiments conducted between 2018 and 2022. The combined statistical and systematic error
on the fit is of the order 5\% for $\pi^+$ and 10\% for $\pi^-$ (applicable
only for the deuteron target). Since the radiative tail from exclusive
pion production varies from 1\% to 10\%, we estimate a model
uncertainty of $0.1-0.5\%$ ($0.2 - 1\%$) for the exclusive
pion radiative tails to $\pi^+$ ($\pi^-$) production. Due to lack
of available data, our simple fit to $\pi \Delta$ production is
much less certain, resulting in a range of $0.5 - 3\%$ uncertainty,
depending on pion kinematics. The ratio of radiated to unradiated
SIDIS cross sections is relatively insensitive to the absolute
normalization of the model and is primarily driven by the kinematic
dependence on $z$ and $P_t$. Based on our iterations of the model,
we estimate about 1\% uncertainty in the radiative corrections
due to the SIDIS model, roughly independent of pion kinematics.

Other sources of radiative correction uncertainty could arise
from the use of the angle-peaking approximation (photons emitted
only along the incident or scattered electron direction), the
uncertainty in the soft-photon term, use of the equivalent
radiator approximation, the 
neglect of pion radiation, and the
lack of two-photon corrections. 

We have listed our results~\cite{webpage} both with and without
radiative corrections, so that future global analyses can improve on the cross section models by incorporating data in kinematic 
regions not constrained by the present experiment, as has been done previously for
DIS data.

\subsection{Relevance to TMD factorization}
The authors of Ref.~\cite{Boglione:2022gpv} have presented three kinematic regions
for pion electroproduction: the target fragmentation region; a soft central region; and 
a current fragmentation region where TMD factorization should be applicable. They 
postulate that the probability of the current fragmentation region
dominating by more than $2\sigma$ requires that the hadron rapidity $y_h<-1$,
corresponding the the regions to the left of the arrows in Fig.~\ref{fig:ptmm0w}. Most
of the present data lie in this region. 

Some previous experiments~\cite{HERMES:2012kpt,COMPASSrho} have subtracted 
contributions from pions originating
from diffractive $\rho$ electroproduction, a process that is clearly not easily
accounted for in the TMD factorization framework. 
Due to a lack of experimental data in our kinematic region, we have not corrected our results 
for the contributions from diffractive vector meson production. Based on an extrapolation of 
the COMPASS collaboration fit~\cite{COMPASSrho}, there could  substantial corrections to the 
$\left<cos(\phi^{*})\right>$ and $\left<\cos(2\phi^{*})\right>$ moments, but small corrections 
for $x>0.1$, where all of the present data lie. 

The data presented in this paper were taken at a single beam energy, and thus cannot
be used to separate the transverse and longitudinal structure functions.

\vspace{-2ex}
\section{Summary}
In summary, we have measured the $\pi^{\pm}$ multiplicities from SIDIS on H and D targets over a three-dimensional grid in  $z$, $P_t$, and $\phi^{*}$ for several
values of $(x,Q^2$). 
The multiplicities were fitted for each bin in $(x,~Q^2,~z,~P_{t})$ with 
three parameters: $\phi^{*}$ -independent $M_0$, and azimuthal modulations
$\langle \cos(\phi^{*}) \rangle$ and $\langle \cos(2\phi^{*}) \rangle$.
The $P_t$-dependence of the $M_0$ results was found to be remarkably consistent 
for the four flavor cases studied: $ep\rightarrow e \pi^+ X$, 
$ep\rightarrow e \pi^- X$, $ed\rightarrow e \pi^+ X$, $ed\rightarrow e \pi^- X$
over the range $0<P_t<0.4$ GeV, as were the multiplicities evaluated near 
$\phi^* = 180^\circ$ over
the extended range $0<P_t<0.7$ GeV. Gaussian widths of the $P_t$-dependence
exhibit a quadratic increase with $z$. 
The $\cos(\phi^{*})$
modulations were found to be 
consistent with zero for $\pi^+$, 
in agreement with previous world data, while the $\pi^-$ moments were in many
cases significantly greater than zero. 
The $\cos(2\phi^{*})$ modulations
were found to be consistent with zero. 
The higher statistical precision of this dataset compared to previously published
data should allow improved determinations of quark transverse momentum distributions 
and higher twist contributions.

The spin-averaged SIDIS results of this paper are but one part of a larger
program at JLab, which includes the use of nuclear targets, polarized
beams and targets, and a range of beam energies to separate
longitudinal and transverse structure functions.

\section{Acknowledgments}

We are grateful to M. Cerutti for providing calculations of the
2022 MAP model at each of the kinematic settings of this experiment

This work was funded in part by the U.S. Department of Energy, including contract 
AC05-06OR23177 under which Jefferson Science Associates, LLC operates Thomas Jefferson National Accelerator Facility, and by the U.S. Department of Energy, Office of Science, contract numbers DE-AC02-06CH11357, DE-FG02-07ER41528, DE-FG02-96ER41003, and by the U.S. National Science Foundation grants PHY 2309976, 2012430, 2013002 and 1714133 and the Natural Sciences and Engineering Research Council of Canada grant SAPIN-2021-00026. We wish to thank the staff of Jefferson Lab for their vital support throughout the experiment. We are also grateful to all granting agencies providing funding support to the authors throughout this project.


\bibliographystyle{apsrev4-1}
\bibliography{main}


\end{document}